\documentclass[twocolumn]{aastex631}

\begin{document}

\title{The effects of the carbon-to-oxygen ratio on the condensate compositions around Solar-like stars}

\author[0000-0003-4709-2689]{Cody J. Shakespeare}
\email{shakec1@unlv.nevada.edu}
\affiliation{Department of Physics and Astronomy, University of Nevada, Las Vegas, 4505 S. Maryland Pkwy, Las Vegas, NV 89154, USA}
\affiliation{Nevada Center for Astrophysics, University of Nevada, Las Vegas, 4505 S. Maryland Pkwy, Las Vegas, NV 89154, USA}

\author{Min Li}
\affiliation{College of Physics, Jilin Normal University, Siping, Jilin 136000, China}
\affiliation{Key Laboratory of Functional Materials Physics and Chemistry of the Ministry of Education, Jilin Normal University, Changchun, Jilin 130103, China}

\author[0000-0001-7660-8766]{Shichun Huang}
\email{shuang25@utk.edu}
\affiliation{Department of Earth and Planetary Sciences, University of Tennessee, Knoxville, 1621 Cumberland Ave, Knoxville, TN 37996, USA}

\author{Zhaohuan Zhu}
\email{zhaohuan.zhu@unlv.edu}
\affiliation{Department of Physics and Astronomy, University of Nevada, Las Vegas, 4505 S. Maryland Pkwy, Las Vegas, NV 89154, USA}
\affiliation{Nevada Center for Astrophysics, University of Nevada, Las Vegas, 4505 S. Maryland Pkwy, Las Vegas, NV 89154, USA}

\author{Jason H. Steffen}
\email{jason.steffen@unlv.edu}
\affiliation{Department of Physics and Astronomy, University of Nevada, Las Vegas, 4505 S. Maryland Pkwy, Las Vegas, NV 89154, USA}
\affiliation{Nevada Center for Astrophysics, University of Nevada, Las Vegas, 4505 S. Maryland Pkwy, Las Vegas, NV 89154, USA}



\begin{abstract}
The initial stellar carbon-to-oxygen (C/O) ratio can have a large impact on the resulting condensed species present in the protoplanetary disk and, hence, the composition of the bodies and planets that form. The observed C/O ratios of stars can vary from 0.1-2. We use a sequential dust condensation model to examine the impact of the C/O ratio on the composition of solids around a Solar-like star. We utilize this model in a focused examination of the impact of varying the initial stellar C/O ratio to isolate the effects of the C/O ratio in the context of Solar-like stars. We describe three different system types in our findings. The Solar system falls into the silicate-dominant, low C/O ratio systems which end at a stellar C/O ratio somewhere between 0.52 and 0.6. At C/O ratios between about 0.6 and 0.9, we have intermediate systems. Intermediate systems show a decrease in silicates while carbides begin to become significant. Carbide-dominant systems begin around a C/O ratio of 0.9. Carbide-dominant systems exhibit high carbide surface densities at inner radii with comparable levels of carbides and silicates at outer radii. Our models show that changes between C/O=0.8 and C/O=1 are more significant than previous studies, that carbon can exceed 80\% of the condensed mass, and that carbon condensation can be significant at radii up to 6 AU.
\end{abstract}

\keywords{Astrochemistry (75), Carbon planets (198), Exoplanet structure (495), Extrasolar rocky planets (511), Planet hosting stars (1242), Stellar abundances (1577)}


\section{Introduction} \label{sec:intro}
The chemical composition of the Sun and its minor bodies help us constrain the Solar system's formation and evolution. Many dust condensation models are able to largely reproduce the composition of the Solar System's rocky bodies. Sequential condensation models reproduce finer details observed in Solar System bodies that more basic models cannot \citep{Li2020}. This makes sequential models more preferred, even more so because the sequential condensation of elements/chemicals over time is more physically motivated than models which assume all species condense at the same time, such as equilibrium condensation models with static disks. The benefits of sequential condensation models especially outweigh their computational cost when examining exoplanet systems around stars with compositions significantly different from the Sun's \citep{Mori2014}. Dust condensation can have fundamental impacts on exoplanet formation: from the system architecture being influenced by the initial surface density of solids to interior structure and surface chemistry. In the absence of data on the composition of exoplanets, we apply an advanced sequential condensation model to predict the impact of elemental ratios on planet composition and formation around stars similar to the Sun. We build upon previous work by examining one of the most important elemental ratios in forming rocky planets: the carbon-to-oxygen ratio.

\subsection{Sequential Condensation}
Chondritic parent bodies did not experience large-scale differentiation processes that melted the planets and formed a layered, core-mantle-crust structure. As such, chondrites provide a record of the physical and chemical processes that occurred early in the Solar nebula and planet-forming disk. Some chondrite types and terrestrial planets are depleted in volatiles relative to the Solar composition -- represented by CI chondrites \citep{Lodders2003}. Volatiles are the elements that begin condensing at relatively low temperatures with refractories condensing at relatively high temperatures. Chemical species similarly form condensates (i.e. solids or minerals, often called dust) at certain temperatures. These species sometimes allow elements to condense at temperatures inconsistent with their elemental condensation temperature. The volatility-refractory behavior of chemical species is a function of pressure and temperature, as well as bulk chemical composition \citep{Li2020}. 

The pressure and temperature (PT) conditions are determined by the evolution of the protoplanetary disk that formed the Solar system. Different areas of the Solar system experienced different cooling histories as the protoplanetary disk formed from the Solar nebula. This uneven cooling leads to the condensation of different elements, and species, at different locations: higher condensation of refractory elements at high temperatures and higher condensation of volatile elements at low temperatures.

The material of the protoplanetary disk is not static either. Pressure gradients cause radial velocities that can move gaseous species, as well as some condensed ones, to different locations. Hydrogen, as the prominent gas, is most affected by these gradients. However, the flow of hydrogen will drag other gaseous species, and some dust, with it. Thus, the bulk chemical composition at a given location in the disk evolves with time, which, in turn, affects the condensation of species and hence elements. This makes evolving disk models an improvement over static models because they capture the sequential and partial condensation of elements over time, which may be especially important in predicting compositions for carbon-rich systems \citep{Mori2014}.

The correlation between the Solar abundances of elements and the composition of the planets and chondrites can be reproduced by a group of popular models that involve partial condensation of volatile elements during chondrite formation \citep{Cassen1996,Petaev2009,Li2020}. Models that show good agreement with the correlation are promising tools for exploring the composition of bodies around stars with different elemental abundances. One important element ratio is the carbon-to-oxygen ratio C/O. Oxygen is a major component in many minerals found in chondrites and terrestrial planets. These include fosterite/olivine (Mg$_2$SiO$_4$), pyroxene (Mg$_2$Si$_2$O$_6$), anorthite (CaAl$_2$Si$_2$O$_8$), diopside (CaMg(SiO$_3$)$_2$), etc. Carbon, on the other hand, is the foundation of many complex organic molecules. Carbon in meteorites is often in the form of graphite, carbonates, and simple organic molecules.

\subsection{Previous Work}

Varying the C/O ratio can have a large impact on the resulting condensed species present in the protoplanetary disk and, hence, the composition of the bodies and planets that form. The observed C/O ratio of stars can vary from 0.1-2 \citep{Bond2010}. The frequency of stars with C/O ratios significantly higher or lower than the observed range is not known. Observations of stars in the Solar neighborhood suggest that this range may be narrower, at least locally, between ratios of 0.1-1 \citep{Brewer2016}. The Sun's C/O ratio of $\sim0.5$ is near the average among the Solar neighborhood \citep{Lodders2003,Brewer2016}. Due to chemical processes, Earth's C/O ratio is estimated at around 0.01 \citep{Allegre2001}. Similarly, at some threshold stellar C/O ratio, often cited as C/O$>0.8$, the bulk C/O ratio of terrestrial planets is expected to be higher than the stellar ratio. This is thought to occur because, with higher C/O ratios, Si begins to primarily condense as SiC, instead of silicate minerals, and the remaining carbon subsequently forms graphite \citep{Bond2010}. Refining the threshold value of the initial stellar C/O ratio requires holding other elements constant while varying the C/O ratio. This restriction limits the scope to Solar-like stars but should increase the certainty of the cause (i.e. the change in the initial C/O ratio) and increase the accuracy of the refined threshold value.

\citet{Bond2010a} attempt to form planet systems similar to the Solar system in orbital architecture and chemical composition. They use a static PT profile from a protoplanetary disk model by \citet{Hersant2001} to correlate the physical locations within the disk to a single point in time where planetesimals condense from the disk. Their chemical compositions are calculated through the commercial equilibrium chemistry software HSC Chemistry. HSC Chemistry takes the elemental abundance of 16 elements and outputs the chemical abundance of 113 species, including 33 condensates, with respect to the PT profile. The chemical composition of the disk is then combined with an N-body planet accretion model by \citet{OBrien2006}. \citet{Bond2010} expand upon \citet{Bond2010a} by applying the same model to known exoplanet systems. \citet{Bond2010} use input compositions derived from the observations of the stellar photosphere of 10 known exoplanet host stars.

\citet{Mori2014} also uses HSC Chemistry to calculate chemical composition. However, they use an evolving disk model from \citet{Chambers2009} that includes both viscous heating and stellar irradiation. \citet{Mori2014} use sequential condensation, similar to \citet{Cassen1996}, which considers the more realistic scenario where solids may condense out of the disk over a range of time. \citet{Mori2014}'s sequential condensation models include the evolving disk's pressure and temperature, the radial motion of gas and coupled dust, and the decoupling of some dust that remains stationary. \citet{Mori2014} use input compositions from the Solar system and three known exoplanet hosts which were also examined by \citet{Bond2010}. They do a small analysis of the impact of the C/O ratio by holding carbon constant while decreasing the amount of oxygen.

\citet{Jorge2022} use a static radial pressure and temperature profile for their equilibrium condensation model that represents the Minimum Mass Solar Nebula from \citet{Hayashi1981}. They use the GGchem code, developed by \citet{Gail1986} and most recently updated by \citet{Woitke2018}, to calculate final chemical compositions. GGchem uses the initial abundance of 24 elements as input with the PT profile to model the formation of 552 species including 241 condensates. \citet{Jorge2022} choose a sample of six G-type stars from the Hypatia Catalog \citep{Hinkel2014} for their initial stellar compositions.

Modern dust condensation models for protoplanetary disks should combine robust chemistry models with disk evolution models. Robust chemistry models have a thorough, but not necessarily exhaustive, number of chemical species and produce results that agree with current data. Disk evolution models enable a robust chemistry model to form chemical species over the range of conditions that will occur sequentially as time passes. The initial elemental composition and the final PT condition, which many static models use, may not be sufficient in predicting the end result of a dynamic condensation history. In reality, solid species sequentially condense out of the disk. Some elements will be depleted earlier preventing condensation at a later time. In addition to element depletion, the PT condition also varies with time. So, a single point in time will not represent the range of PT values experienced at a particular location. Furthermore, gas can flow radially in the protoplanetary disk and the flow can change direction and strength over time. This gas flow can move light elements to other parts of the disk where they may then condense, enriching some areas with volatiles and depleting others. The flowing gas may also carry some advected dust with it, displacing solid species from their initial formation location. These are factors that our model takes into account. Note, however, that the previously mentioned studies, and our model here, do not include volatile depletion effects that may occur after condensation. Volatile depletion after condensation is considered a second-order effect and it is not needed to reproduce the major chemical trends of the solar system (e.g. \citet{Bond2010}, \citet{Mori2014}, \citet{Li2020}, \citet{Jorge2022}).

Our examination of the C/O ratio is an improvement over many others in three important aspects. First, we use a robust chemistry model, GRAINS \citep{Petaev2009}, that includes more elements, chemical species, and condensates than recent studies. GRAINS is a chemical condensation code that uses 33 elements with 762 chemical species including 520 condensates, or solid dust species, and 242 gaseous species. Second, we combine this modern chemistry model with an evolving disk model from \citet{Cassen1996} to conduct sequential condensation chemistry \citep{Li2020}, similar to the sequential condensation in \citet{Mori2014}. Third, we utilize this model in a focused examination of the impact of C/O ratios to provide robust results and conclusions in the context of Solar-like stars. Our focused examination explores a range of elemental ratios while holding all other elements constant with Solar values. Using stellar abundances makes it difficult to isolate the effects of a specific elemental ratio as other ratios vary simultaneously. Varying ratios over a range allow a close examination of the effect that small differences in elemental ratios have on condensate composition -- and eventually planets -- rather than relying on samples of stellar abundances.

\section{Model and Methods}

\begin{figure}
    \centering
	\includegraphics[width=\columnwidth]{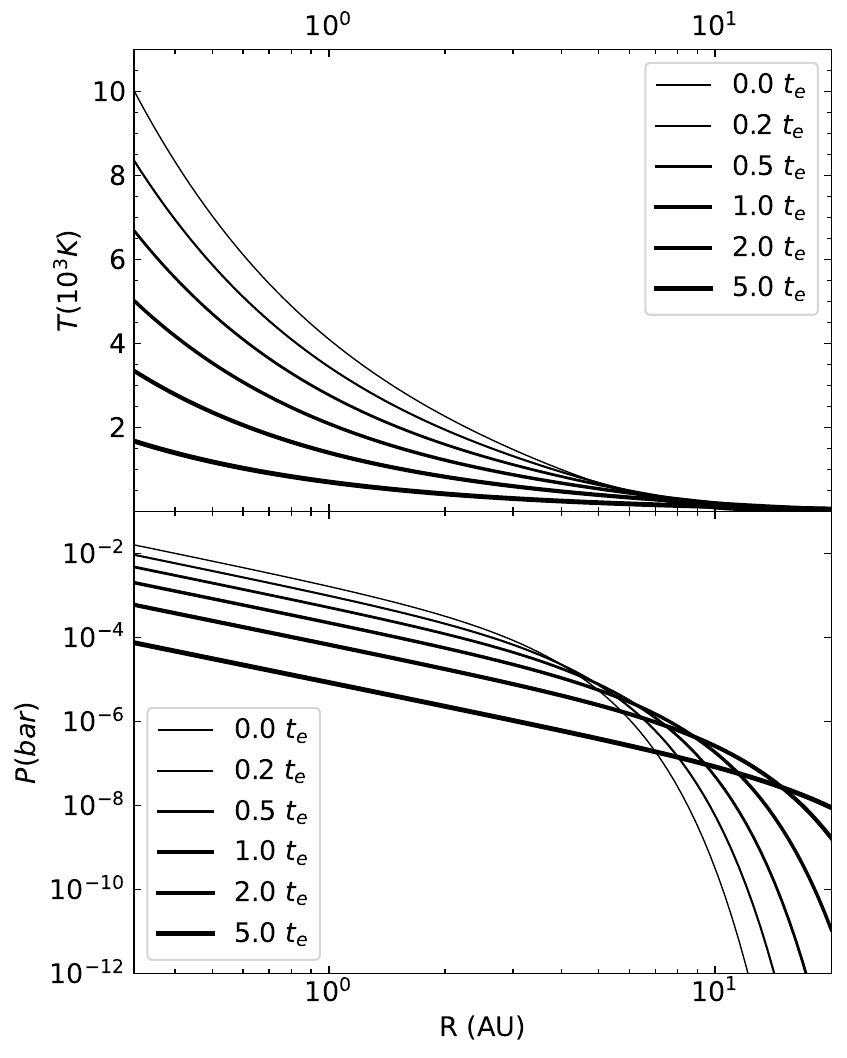}
    \caption{The evolution of the gaseous disk parameters from the beginning (0.0 $t_e$) to the end (5.0 $t_e$) of our simulations. (Top) Midplane temperature (Bottom) Midplane pressure}
    \label{fig:PT}
\end{figure}

\subsection{Model}
\label{sec:Model}

The disk evolution model in \citet{Cassen1996} was combined with the chemical condensation code GRAINS \citep{Petaev2009} in the previous work \citet{Li2020}.  In this study, we use the same model because, using Solar values, its results agree with measured chondritic and planetary compositions \citep{Li2020}. We use this combined sequential condensation with disk evolution code to explore the condensation of species around the early Sun while varying select elemental ratios.

Our algorithm partitions elements in three phases. One gas phase and two phases that are condensates: advected dust and decoupled dust. Advected dust flows with the gaseous phase and may continue to chemically interact with its environment. Decoupled dust remains stationary and chemically inert. The gaseous phase consists of 242 chemical species which react chemically depending on the local conditions. The gaseous and advected species flow with the same velocity as the H$_2$ and He. Depending on the local conditions, the gaseous species may react chemically and subsequently condense into advected or decoupled dust. Simpler dust species may form from a simple phase change instead.
\begin{table*}[]
	\caption{The ratios explored in our study. Each number corresponds to an individual simulation with the listed value set as the corresponding ratio (C/O or N/O). C/O w/Cc simulations were run at C/O ratios that correspond to certain N/O ratios, denoted by their respective superscript. Note that for N/O=0.786 and N/O=0.501 w/Oc, the amount of carbon and nitrogen are equal. The superscripts $a$ and $b$ denote the C/O w/Cc simulations which have the same C/O ratio as the corresponding N/O simulations.}
	\label{tab:sims}
	\centering
        \small
        \tabcolsep=0.09cm
    \begin{tabular}{lcccccccccccccccccccccc}
		\hline
		\hline
		C/O & 0.01 & 0.1 & 0.25 & 0.333 & 0.416 & 0.5 & 0.52 & 0.6 & 0.7 & 0.75 & 0.8 & 0.9 & 0.98 & 1 & 1.1 & 1.2 & 1.3 & 1.5 & 2 & - & 10 & 100\\
		C/O w/Oc & 0.01 & 0.1 & 0.25 & - & - & - & - & - & 0.7 & - & - & 0.9 & - & 1 & 1.1 & - & - & 1.5 & 2 & 4 & 10 & 100\\
        C/O w/Cc & - & - & - & - & - & - & - & - & 0.7 & - & - & - & - & 1 & 1.1$^a$ & - & - & 4.84$^b$\\
		\hline
		N/O & 0.0138 & 0.05 & 0.1 & 0.125 & 0.166 & 0.25 & 0.333 & 0.5 & - & - & 0.75 & 0.786 & - & - & 1.5$^a$ & 1.75 & 2 & 10$^b$\\
		N/O w/Oc & 0.0138 & - & 0.1 & 0.125 & 0.166 & 0.25 & 0.333 & 0.5 & 0.501 & 0.7 & 0.75 & - & 0.8 & 1 & - & - & 2 & 10\\
		\hline
		\hline
	\end{tabular}
\end{table*}
Our gaseous disk evolves according to the surface density profile
\begin{equation}
    \Sigma(r,t)=\Sigma_0(t)e^{-[r/r_0(t)]^2}.
    \label{eq:Sigmat}
\end{equation}
$\Sigma_0(t)$ is the gaseous surface density near the star given by
\begin{equation}
    \Sigma_0(t)=\frac{M_d}{\pi r_0^2},
    \label{eq:Sigma0}
\end{equation}
where $r_0$ is the characteristic radius
\begin{equation}
    r_0(t)=\frac{1}{GM_*}\left(\frac{J}{M_d \Gamma(5/4)} \right)^2.
    \label{eq:r0}
\end{equation}
Here $M_d$ is the disk mass, $G$ is the gravitational constant, $M_*$ is the mass of the star, $J$ is the angular momentum of the disk, and $\Gamma$ is the Gamma function.

The disk mass changes with time as
\begin{equation}
    M_d(t)=M_{d0}\left(1+\frac{t}{te}\right)^{-0.5}.
    \label{eq:Mdt}
\end{equation}
The initial disk mass, $M_{d0}$, is $0.21M_\odot$ and our characteristic evolution timescale $t_e$ is $2.625 \times 10^4$ yrs using our parameters.

The disk is heated by the viscous accretion of gas. The midplane temperature is calculated at each radius and time using
\begin{equation}
    T^4=\frac{3G\tau M_* \dot{M_I}}{64\pi \sigma_{SB}r^3}.
    \label{eq:Tc}
\end{equation}
Here $\dot{M_I}$ is the accretion rate of gas onto the star, $\sigma_{SB}$ is the Stefan-Boltzmann constant, and the optical depth is $\tau=\kappa \Sigma/2$. The temperature and pressure evolution are shown in Figure \ref{fig:PT} for the duration of the simulation. 

\citet{Li2020} found good agreement between observed elemental abundances in chondrites and our model out to 4 AU. Beyond 6 AU, parts of the disk can spend most of the simulation time below 300 Kelvin where highly volatile species (e.g. H$_2$O, CO$_2$, simple hydrocarbons, and others) become important. Our model includes these species in the gaseous phase, but stops calculating condensation below 300 K as it is intended to examine the rocky planets and planetesimals in the inner Solar system. This means we do not fully model the condensation of volatile species. Although some volatile condensates, like Ice I, are included in our model, the abundance of Ice I is very low above 300 K.

At a time of 2 $t_e$, 2/5 of the total simulation time, 6.64 AU is the innermost radius that has cooled to 300 K (see Figure \ref{fig:PT}). The results of our model do not correctly represent the condensation of highly volatile species in areas outside of 3.1 AU which cool below 300 K before the end of our simulation. Furthermore, the results of our model may not properly represent the condensation of refractory species beyond 6 AU due to the large fraction of time falling below the 300 K limit. In other words, at outer radii, especially beyond 6 AU, refractory species condensation in our model may not be a reliable representation of reality. Any figure that goes beyond 6 AU includes a shaded region in the outer area to emphasize the high uncertainty.

A more in-depth discussion of the model used can be found in \citet{Li2020}, but also in \citet{Petaev2009} and \citet{Cassen1996}. The model results using Solar values are compared against chondritic and planetary compositions in \citet{Li2020}.

\subsection{Varying Ratios}
\label{sec:Ratios}

We vary the initial conditions (i.e. the initial C/O and N/O ratio) to examine the impact on the final properties of the disk. There are many different ways to vary the C/O and N/O ratios. We search the C/O and N/O parameter space with the following two methods. The first method keeps the initial sum of the number of atoms for the two elements constant while the ratio is varied. For example, when varying the C/O ratio, we keep the sum of $N(\text{C})$ and $N(\text{O})$ constant while varying the ratio
\begin{equation}
    \text{C/O}=\frac{N(\text{C})}{N(\text{O})},
    \label{eq:ratio}
\end{equation}
where $N(\text{E})$ is the number of atoms of element ``E'' per 1 million Si atoms. Simulations using this method are simply referred to as C/O ratios, or as “with constant sum” in discussions that may be confusing without a distinction.

In the second method, we vary the ratios while holding $N(\text{O})$ constant at the Solar value. Thus, only $N(\text{C})$ is varied to achieve different ratios. We denote these simulations ``C/O w/Oc'' as shorthand for the C/O ratio with constant oxygen. Holding oxygen constant for some simulations helps us to determine if having a low relative abundance of oxygen, compared to carbon, has differing effects compared to a low absolute abundance of oxygen. In other words, having two methods distinguishes between the effects caused by varying ratios and the effects caused by oxygen depletion, the latter of which occurs in high C/O ratios with a constant sum.

The first method, ``constant sum'', benefits from having no additional atoms added to the simulation. This is useful when comparing the fraction of condensed mass attributed to carbon and surface densities. The former will converge to 100\% when large amounts of carbon are added in the ``w/Oc'' method as carbon will eventually be the most common element. In other words, the carbon abundance in the C/O w/Oc method is unbounded as the C/O ratio goes to infinity. Surface densities will be easier to compare for the ``constant sum'' method because the mean mass of the atoms will be nearly the same; only varying because of the 4 AMU difference between carbon-12 and oxygen-16. The ``constant sum'' method will also benefit by examining systems that have a higher $N(\text{O})$ value, up to about the Solar sum of $N(\text{C})$ and $N(\text{O})$. The second method, ``w/Oc'', benefits by keeping the ratio of all other elements to oxygen constant (i.e. $N(\text{E})/N(\text{O})$ remains at the Solar value), thus, helping to isolate the cause to the change in $N(\text{C})$.

We search a parameter space of C/O and N/O ratios to find transitions in the behavior of final species compositions and distributions. The ratios simulated are shown in Table \ref{tab:sims}. Only one ratio is varied in each simulation (i.e. the C/O and N/O ratio are not varied simultaneously). Due to the nature of varying $N(\text{O})$ when having a constant sum, other ratios with oxygen are indirectly varied. The Solar ratio simulation, first explored in \citet{Li2020}, corresponds to C/O=0.5010 and N/O=0.1380 \citep{Lodders2003} and is not listed in Table \ref{tab:sims}. We examine smaller changes of ratios in areas of sharp compositional change (see Table \ref{tab:sims} and Section \ref{sec:Results}) or examine ratios consistent with observations and predictions of stellar populations \citep{Bond2010}.

We aim to refine the threshold C/O ratio, often cited at around C/O=0.8, and determine if there is a similar threshold N/O ratio where condensate composition changes. We also aim to better understand how elemental ratios change the distribution of common condensed species that may form terrestrial planets. These threshold ratios and composition changes also depend on star mass and other parameters like disk mass, stellar rotation rate, etc. We limit our scope to Solar-like stars for this reason, but our results likely apply to many star types to varying degrees. Similar to \citet{Bond2010}, we attempt to categorize final compositions as “silicate-dominant”, where oxygen is abundant, or “carbide-dominant”, where possible, by comparing them to extreme ratios. Note that silicate-dominant, or oxygen-dominant, is largely synonymous with oxidized environments and carbide-dominant, or carbon-dominant, is largely synonymous with reduced environments.

\subsection{Methods of Analysis}
\label{sec:Figures}

One common method of examining elemental ratio results is the $R_i$ in Equation \ref{eq:Ri} below. This is the ratio of the concentration of an element, $\chi_E$, to the concentration of silicon, $\chi_{Si}$, for a certain planet or chondrite, denoted by the subscript $\textbf{P}$. This ratio is then normalized to the same ratio for the Sun in the denominator of Equation \ref{eq:Ri}. \citet{Li2020} compare the $R_i$ of our model results at certain distances from the Sun to the measured $R_i$ of chondrites. In this paper, we do not include the data points of chondrites, so the subscript $\textbf{P}$ solely refers to our modeled condensates at certain distances from the Sun. Our $R_i$ results will be presented in Figures \ref{fig:LiFig9CO} and \ref{fig:LiFig9NO}. See Fig. 9 of \citet{Li2020} for the comparison with the chondritic data points.
\begin{equation}
    R_i=\frac{[\chi_E/\chi_{Si}]_\textbf{P}}{[\chi_E/\chi_{Si}]_\odot}
    \label{eq:Ri}
\end{equation}

Providing a detailed breakdown of the molecular species in a similar manner is difficult due to the sheer number of species we simulate. We limit the included species to the most significant ones using the following method. The species were filtered by first arranging them from greatest to least, in terms of their condensed fraction, and then summing the species until a total condensed fraction of 90\% was achieved. This was done at the radii of 0.5, 1, 2, and 4 AU for each simulation listed. A species was removed if it does not contain carbon, oxygen, or nitrogen, or does not surpass 5\% of the total mass in any simulation or radius. Some species are exceptions to this rule as they appear to be good markers for silicate-dominant and carbide-dominant systems. We will present the maximum and minimum mass fraction of the significant species at these four radii, in addition to marking the value at 1 AU, in our Figures \ref{fig:SpecCnameEXTR}-\ref{fig:SpecNname}.

\section{Results}
\label{sec:Results}

We use the final C/O ratio and the fraction of total mass that is attributed to either oxygen or carbon to examine the impact of elemental ratios from a broad perspective in Section \ref{sec:Elements}. For a more detailed perspective, we examine the elemental surface profile of condensates for select simulations. We also examine the impact of elemental ratios on the $R_i$ (see Eq. \ref{eq:Ri}) of many elements. In Section \ref{sec:Species}, we attempt to categorize systems into either carbide-dominant or silicate-dominant archetypes based on the composition of condensed species. Examining the species most responsible for changing elemental ratios helps with understanding results and providing metrics to further improve our model.
\begin{figure*}
	\includegraphics[width=0.49\textwidth]{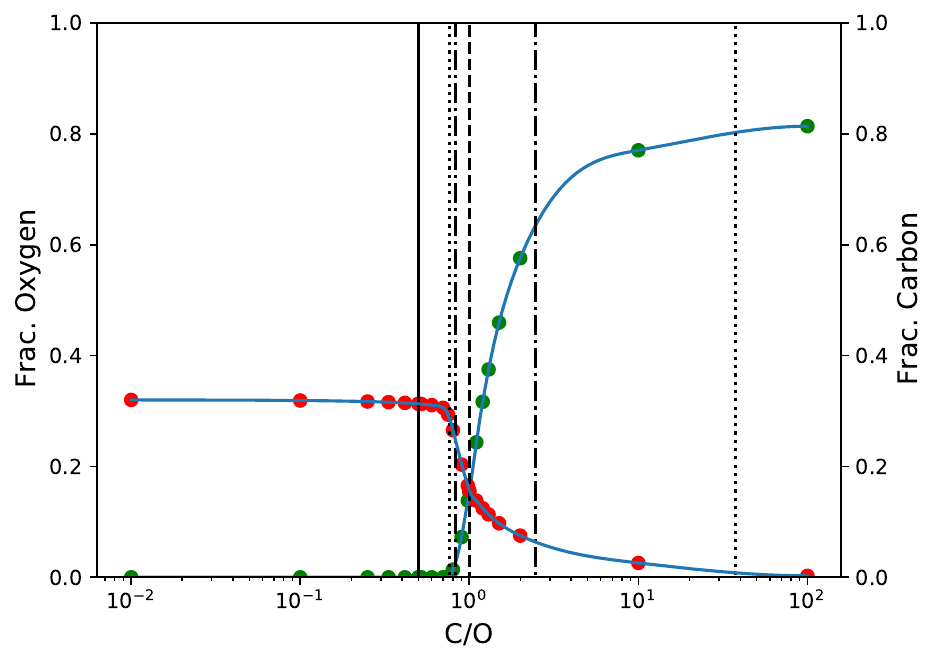}
	\includegraphics[width=0.49\textwidth]{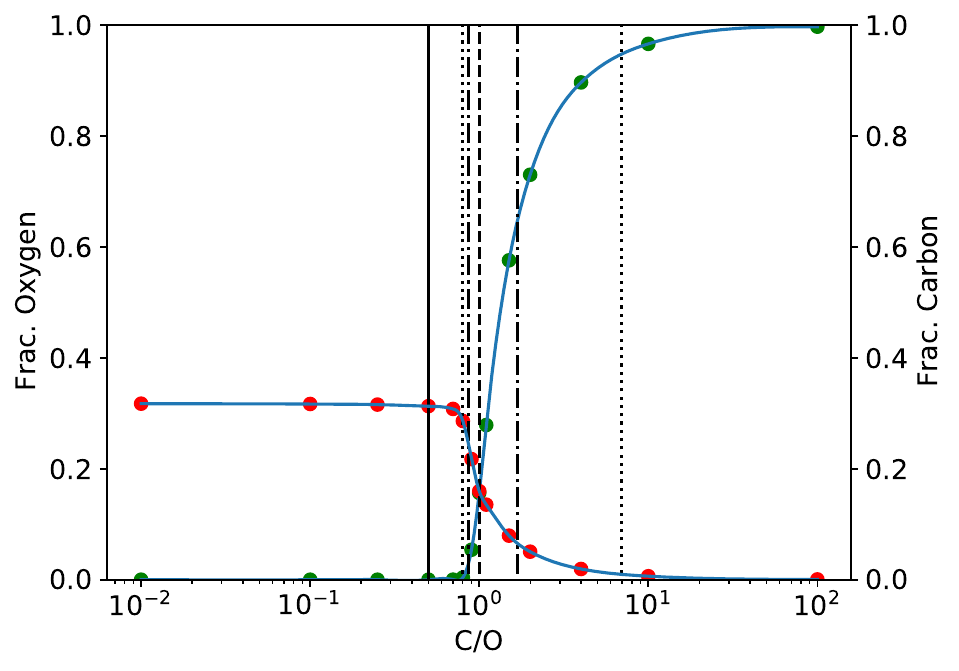}
    \caption{The total carbon (green) and oxygen (red) fractions summed over all radii for simulations with a given initial C/O ratio. The vertical solid line indicates the Solar C/O ratio simulation. Purely dotted lines correspond to $\frac{f\text{C}}{f\text{O}}=0.01$ and $100$. Dashed and dotted lines correspond to $\frac{f\text{C}}{f\text{O}}=0.1$ and $10$. The purely dashed line corresponds to $\frac{f\text{C}}{f\text{O}}=1$. (Left) In the order from left-to-right: $\frac{f\text{C}}{f\text{O}}=0.01$, $\frac{f\text{C}}{f\text{O}}=0.1$, $\frac{f\text{C}}{f\text{O}}=1$, $\frac{f\text{C}}{f\text{O}}=10$, $\frac{f\text{C}}{f\text{O}}=100$ are located at C/O values of 0.762, 0.828, 1.005, 2.457, and 37.633, respectively. (Right) Here the amount of oxygen is held constant at the Solar value while the C/O ratio varies (i.e. C/O w/Oc). In the order from left-to-right: $\frac{f\text{C}}{f\text{O}}=0.01$, $\frac{f\text{C}}{f\text{O}}=0.1$, $\frac{f\text{C}}{f\text{O}}=1$, $\frac{f\text{C}}{f\text{O}}=10$, $\frac{f\text{C}}{f\text{O}}=100$ are located at C/O values of 0.798, 0.859, 1.005, 1.684, and 6.964, respectively. }
    \label{fig:FracLine}
\end{figure*}
\begin{figure*}
    \centering
	\includegraphics[width=0.48\textwidth]{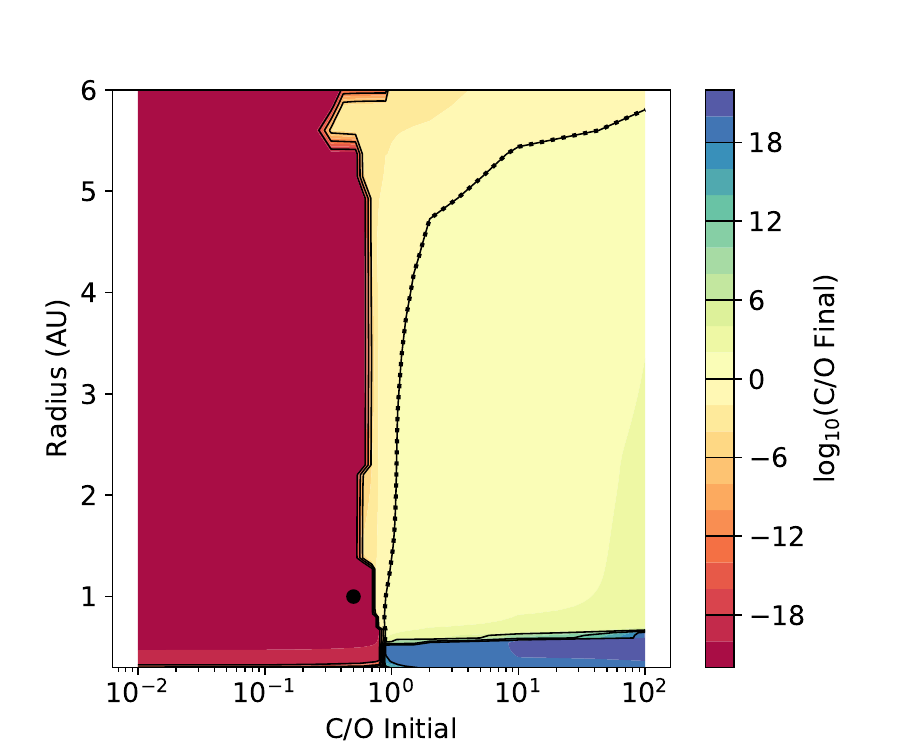}
 	\includegraphics[width=0.48\textwidth]{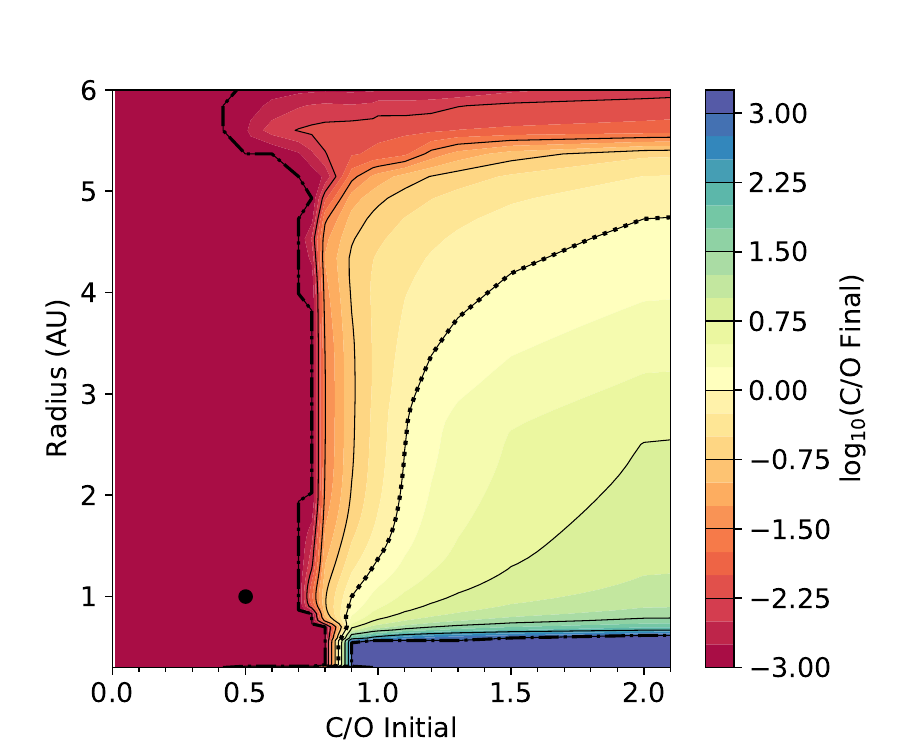}
    \caption{ The final C/O ratio, denoted by color, at different distances from the central star and initial C/O ratios. The y-axis is a linear scale of the radii in AUs. The z-axis, or colored contours, is a log scale of the final local C/O ratio for the given simulation's initial conditions and at the indicated distance from the central star. Black lines help emphasize each change of $10^6$ and the contour line for C/O=1 is represented with a dotted line. The black dot located at C/O=0.501 and R=1 AU represents the location and initial C/O ratio of Earth's position in the Solar protoplanetary disk. Note that the resolution of these two contour plots is not uniform between the x and y-axis. The resolution is the poorest at the far left, far right, and top. Lastly, note that the resolution of the x-axis is not uniform, see table \ref{tab:sims} for the exact initial C/O ratios we explore. (Left) The x-axis is a log scale of the initial C/O ratio spanning all ratios examined. (Right) The x-axis is a linear scale and spans C/O ratios of 0-2.1. The z-axis here is limited from 10$^{-3}$-to-10$^3$ to better display small changes in the final C/O ratio. The dash-dotted lines in the right plot represent the 10$^{-3}$ and 10$^3$ cutoffs in the z-axis.}
    \label{fig:Frac3D}
\end{figure*}
\subsection{Elemental Composition}
\label{sec:Elements}

The abundance of elements in the solid dust is also impacted by increasing C/O ratios due to the preferential binding by carbon or oxygen. For example, a depletion in one reactant, oxygen, can be caused by preferential bonding to carbon in the gaseous CO and CO$_2$. With the oxygen preoccupied by carbon, other elements we typically find with oxygen, such as metals, alkali metals, silicon, etc., may experience an increase in abundance in their metallic and/or carbide form. Thus, allowing the formation of condensates we would not find in an oxygen-dominant system like the Sun. This is only one illustrative example that is occurring simultaneously with all the other elements and species used in our models. Other element-species pairs, like silicon being preoccupied in SiC, may be more important \citep{Bond2010} depending on pressure, temperature, and local abundances in the disk.

This section will provide an analysis of the results from the perspective of the elements in the solid phase. The elements contained in the solid dust hold particular importance because once this dust is incorporated into planetesimals, the elemental composition is difficult to change. The chemical species is subject to change from chemical reactions during heating events like accretion, impacts, etc., but the elements will generally remain inside the planetesimal.

The following section consists of three parts. The first, section \ref{sec:Threshold}, will refine the threshold C/O ratio for a Solar-like protoplanetary disk. The second, Section \ref{sec:Ele2}, will cover where certain elements condense with respect to radius, or distance from the Sun. The third, Section \ref{sec:Ele3}, will cover how the elemental composition of the condensed dust change with respect to the variation of initial elemental ratios.

\subsubsection{Refining the Threshold C/O Ratio}
\label{sec:Threshold}

One of our objectives is to better measure the threshold C/O ratio where the bulk composition of solids changes significantly, often cited as approximately C/O=0.8 \citep{Bond2010}. Figure \ref{fig:FracLine} shows the final mass fraction of total condensed material that is attributed to oxygen or carbon atoms. The total condensed material is the sum of condensates over all radii. In other words, $f\text{C}=\frac{\Sigma (\text{C})}{\Sigma (\text{All})}$. Summing over all radii allows us to classify a specific system as a whole and avoid local changes at certain radii. The next section will examine how varying the initial C/O ratio changes the local C/O ratios at all radii. Figure \ref{fig:FracLine} has dashed and/or dotted lines to denote when the ratio of carbon mass fraction to oxygen mass fraction, $\frac{f\text{C}}{f\text{O}}$, reaches 0.01, 0.1, 1, 10, and 100. These lines are always ordered from left-to-right, respectively, as high oxygen fractions are located on the left to correspond to the left y-axis.

The exact location of the transition is subjective so we provide the five listed values seen in Figure \ref{fig:FracLine}. The $\frac{f\text{C}}{f\text{O}}=0.01$ and $0.1$ lines best represent the threshold C/O ratio where the behavior begins to change and they are consistent with a value around 0.8. However, these values may change if other elemental ratios, stellar parameters, or disk parameters are varied. In the left panel of Figure \ref{fig:FracLine}, the transition points for $\frac{f\text{C}}{f\text{O}}=0.01$ and $0.1$ occur at C/O ratios of 0.762 and 0.828. The area between $\frac{f\text{C}}{f\text{O}}=0.01$ and $\frac{f\text{C}}{f\text{O}}=1$ is the area with the sharpest change in oxygen mass fraction. For the C/O w/Oc simulations, the right panel of Figure \ref{fig:FracLine} shows the transition points for $\frac{f\text{C}}{f\text{O}}=0.01$ and $0.1$ occur at C/O ratios of 0.798 and 0.859. Most stars in the Solar neighborhood have C/O ratios between approximately 0.1-1 with a median of 0.47 \citep{Brewer2016}. More broadly, stars can have C/O ratios between 0.1-2 with a median of 0.68 \citep{Bond2010}. Thus, systems with C/O ratios between the value at $\frac{f\text{C}}{f\text{O}}=0.01$ or $0.1$ and C/O=1 are the most common systems that will exhibit significantly different compositions from the Solar system.

\subsubsection{Element Distribution over Radius}
\label{sec:Ele2}

The local C/O ratio and the behavior of other elements can vary over radius within a single simulated system. This part of Section \ref{sec:Elements} will first examine how the final, local C/O ratio of the dust varies over radius (i.e. the distance from the central star). Then we will examine how the condensation of other important elements varies over radius. While a system may be broadly classified as carbon or carbide-dominant from the initial C/O ratio, the local variation of the C/O may be important for understanding and predicting the observations of protoplanetary disks, forming planets with varied compositions within the same stellar system, and mixing of elements and species later in the disk's history. Overall, our results do not show significant differences among the methods of varying the C/O ratio. Further figures of the C/O w/Oc or C/O w/Cc results will be in Appendix \ref{sec:Appendix} for comparison with the C/O with ``constant sum'' results in the main text below.

The local C/O ratio at the end of the simulations is shown in Figure \ref{fig:Frac3D} as colored contours. The C/O w/Oc results, which are very similar, are shown in the appendix. Figure \ref{fig:Frac3D} shows a sharp change in composition at initial C/O ratios higher than the Sun's. However, the final, local C/O ratio can differ within a system. Systems with an initial C/O ratio slightly below the Sun's have a carbide-dominant area between 5 and 6 AU in Figure \ref{fig:Frac3D}. At radii smaller than 5 AU, the threshold C/O varies between initial C/O ratios of about 0.6-0.8 depending on the location within the disk. This variation with radius causes intermediate systems -- systems that show silicate-dominant chemistry with reduced surface density and a few percent of carbon mass fraction. Intermediate systems exist in a narrow transition region that begins just above an initial C/O of 0.52 -- the highest C/O simulation that looks nearly identical to the Solar simulation. The transition region ends just above an initial C/O of 0.8. Increasing the C/O ratio further above 0.8 continues to increase the local, final C/O ratios but once C/O=0.9, the behavior is firmly in the carbide-dominant regime where carbides surpass, or nearly match, silicate levels in all regions. Note that we do not fully model the condensation of CO, CO$_2$, hydrocarbons, or H$_2$O due to our chemistry model's lower limit of 300 K. The addition of more volatile species may cause significant changes at outer radii so we limit these figures to radii inside of 6 AU.

Carbide-dominant systems that exceed C/O=1 show carbon surpassing all other elements at nearly all radii inside of 6 AU. Carbide-dominant systems at the low end of C/O ratios, between C/O$\sim$0.9 and C/O=1, show carbon being a dominant element at radii inwards of about 1 AU. Outside of 1 AU, oxygen is more abundant than carbon but carbon remains a major element -- above or near 10\% mass fraction. Low-end carbide-dominant systems may become well-mixed over time leading to planets with a rich spectrum of shared carbide-silicate compositions. Some poorly mixed carbide-dominant systems might retain the local C/O ratios shown in Figure \ref{fig:Frac3D}. These poorly mixed carbide-dominant systems could form more silicate-dominant planets in some regions with carbide-dominant planets forming in other regions. The radius dependence of the C/O ratio is unlike features such as ice lines because the minerals formed in high C/O regions will not evaporate if they move moderately closer to the star. The minerals responsible for the low C/O and high C/O regions in Figure \ref{fig:Frac3D} exist as solids at similar temperatures. However, which mineral forms preferentially from the cooling gas depends on temperature, among other factors, in the equilibrium chemistry calculations.

Our results suggest that low-end carbide-dominant systems that exhibit qualities of both regimes could be common and exhibit varied behavior that depends on mixing. Some past studies have struggled to produce planets with a varied composition of carbides and silicates, as opposed to being mostly one or the other. For example, \citet{Bond2010} only presented the formation of two such mixed-composition planets in their study. \citet{Bond2010}'s Fig. 12 shows the planet in their `Sim.4' of 55 Cnc contains a small but significant fraction of carbon. \citet{Bond2010}'s Fig. 14 shows that the outermost planet in `Sim.3' of HD19994 has a large fraction of carbon and oxygen. However, the mixed composition of \citet{Bond2010}'s HD19994 is more consistent with a carbide-dominant system like our C/O=1 simulation, discussed further below. \citet{Mori2014} produces mixed-composition planets but predicts that carbide-dominant planets will only be able to form at radii inside of 0.3 AU, even for C/O ratios above C/O=1.
\begin{figure*}
	\includegraphics[width=0.96\textwidth]{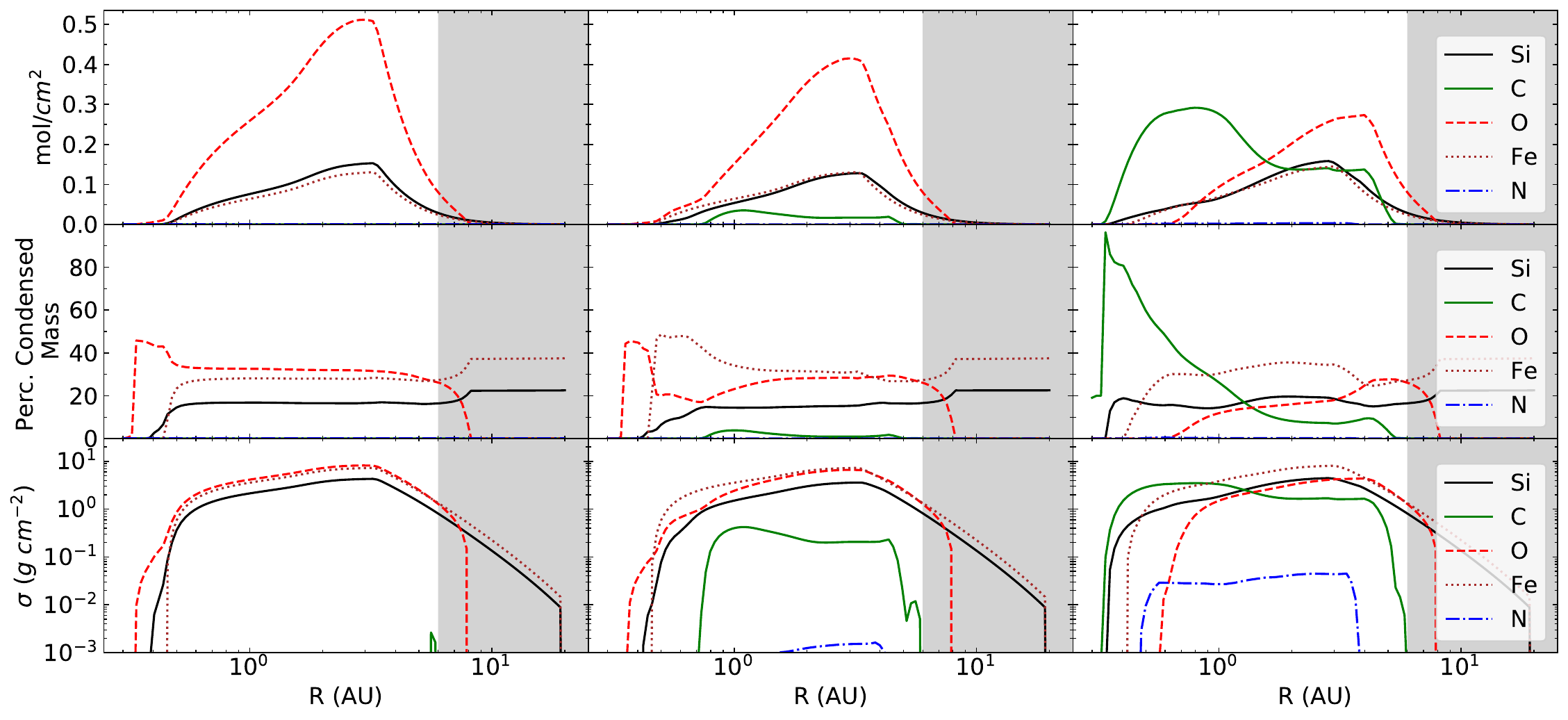}
    \caption{The solid surface density profiles of five select elements in three different simulations: (Left) Solar, (Middle) C/O=0.8, and (Right) C/O=1. The Solar C/O ratio is 0.501 and the Solar N/O ratio is 0.138. The surface density profile does not change significantly between the C/O=1 simulation, the C/O=1 w/Oc simulation, and the N/O=1.5 simulation which has an indirect C/O ratio of 1.10.}
    \label{fig:EleSig1}
\end{figure*}

\begin{figure*}
	\includegraphics[width=0.96\textwidth]{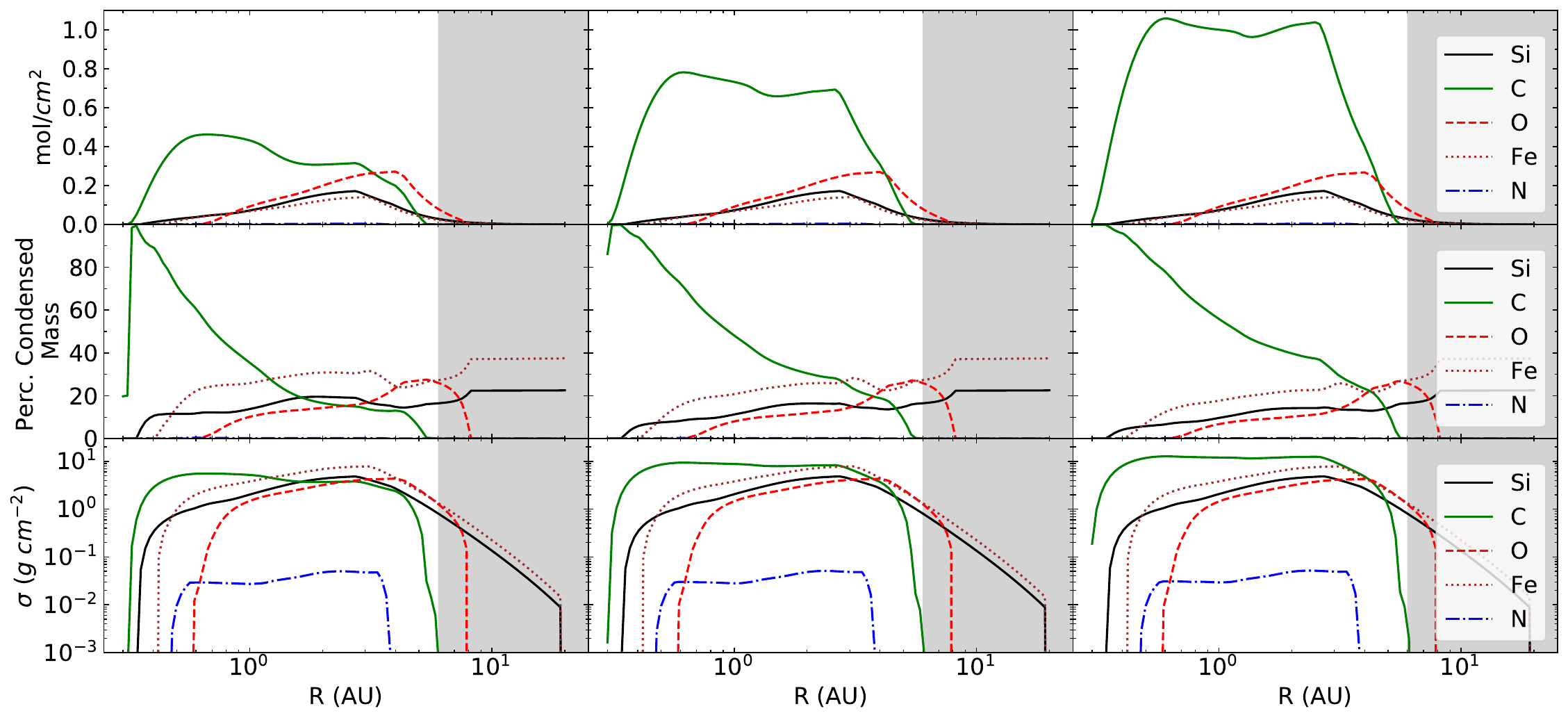}
    \caption{The solid surface density profiles of five select elements in three different simulations: (Left) C/O=1.1, (Middle) C/O=1.3, and (Right) C/O=1.5. Note the y-axis for the molar plot has changed compared to Figure \ref{fig:EleSig1}. The surface density profile of Si, O, Fe, and N does not change significantly between the C/O=1 simulation (see Figure \ref{fig:EleSig1}) and the simulations shown here.}
    \label{fig:EleSigMore}
\end{figure*}
\begin{figure*}
    \centering
	\includegraphics[width=0.34\textwidth]{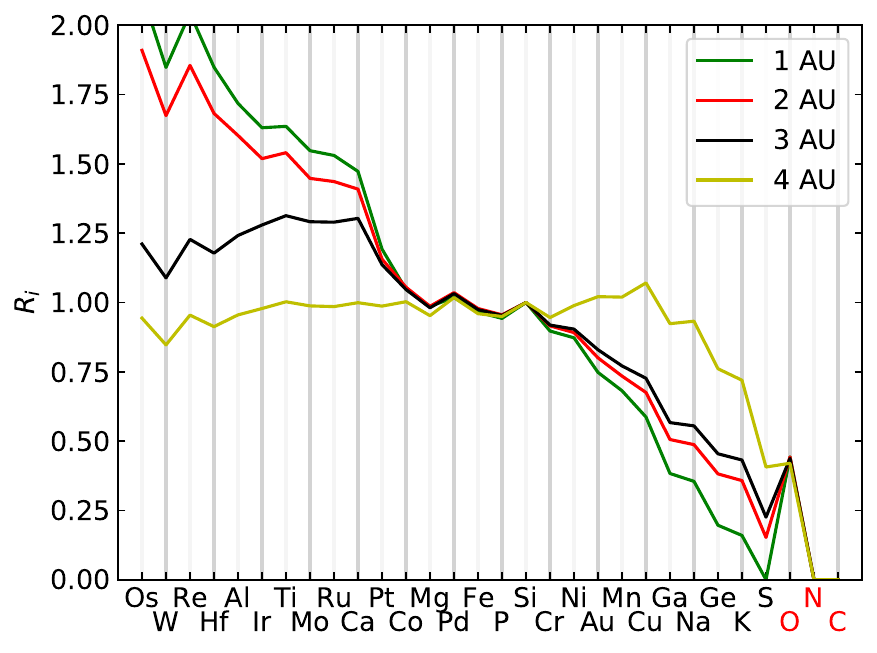}
	\includegraphics[width=0.34\textwidth]{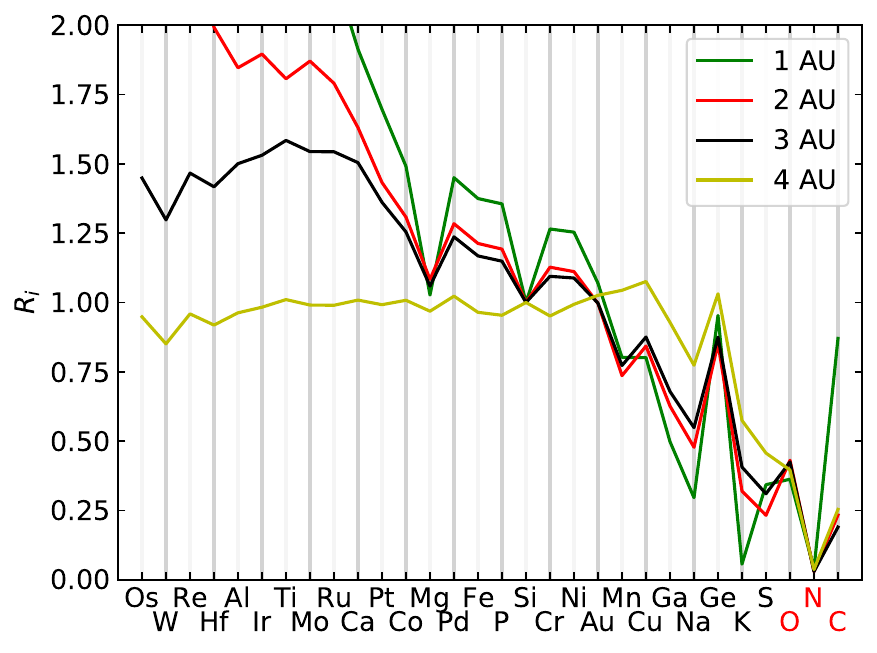}
	\includegraphics[width=0.34\textwidth]{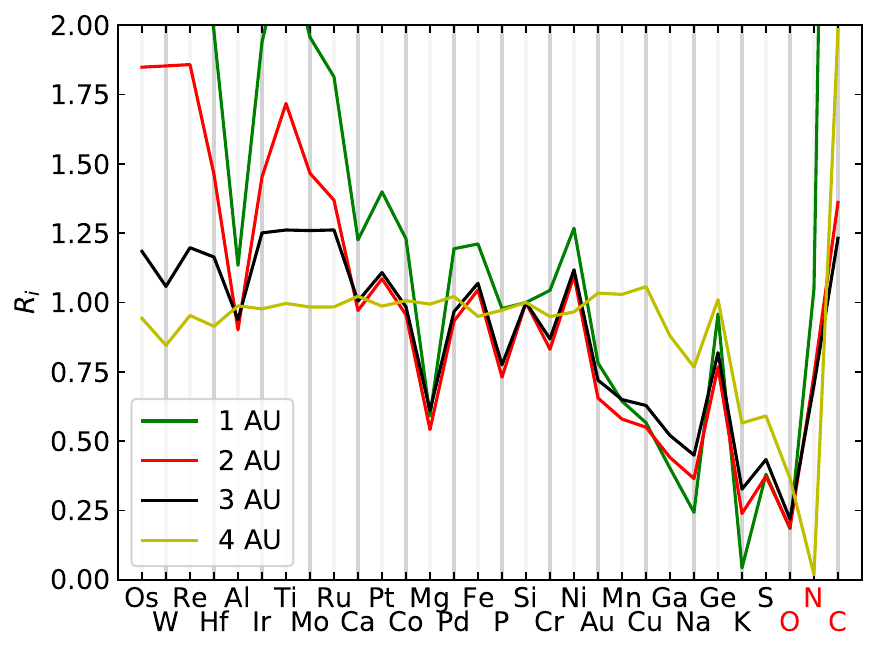}
	\includegraphics[width=0.34\textwidth]{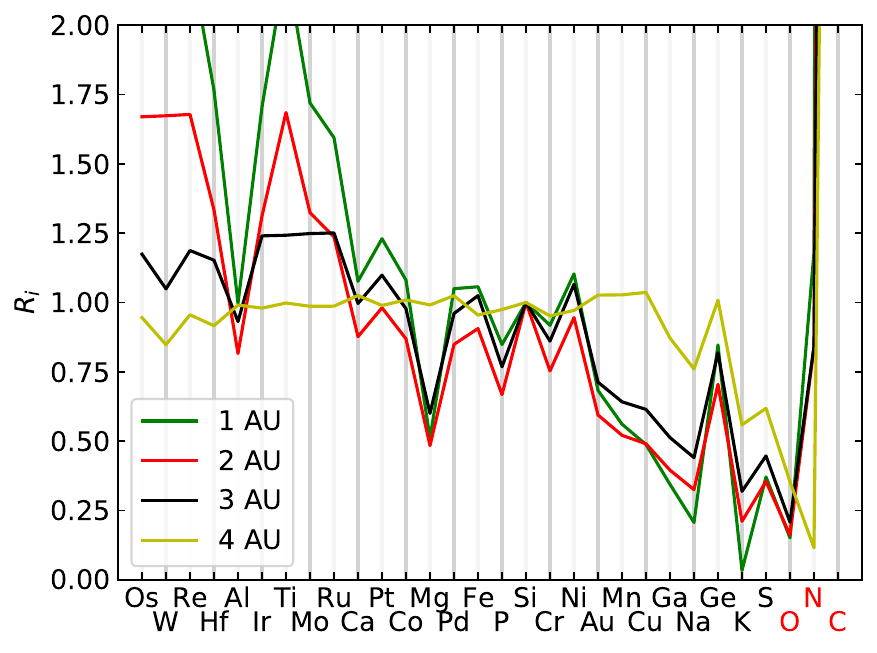}
    \caption{The same figure type as \citet{Li2020}'s Fig. 9 with four different C/O ratios. Carbon, nitrogen, and oxygen are appended to the right with red letters to denote the elements being artificially altered. The top left graph is the same simulation as \citet{Li2020}'s M2G1 model in Fig. 9 and has a Solar C/O ratio which is 0.501. The top right, bottom left, and bottom right correspond to the initial C/O ratios of 0.8, 1, and 2, respectively. Those three simulations are the closest to the $\frac{f\text{C}}{f\text{O}}$=0.1, $\frac{f\text{C}}{f\text{O}}$=1, and $\frac{f\text{C}}{f\text{O}}$=10 lines in Figure \ref{fig:FracLine}, in the same order.}
    \label{fig:LiFig9CO}
\end{figure*}

\begin{figure*}
    \centering
	\includegraphics[width=0.34\textwidth]{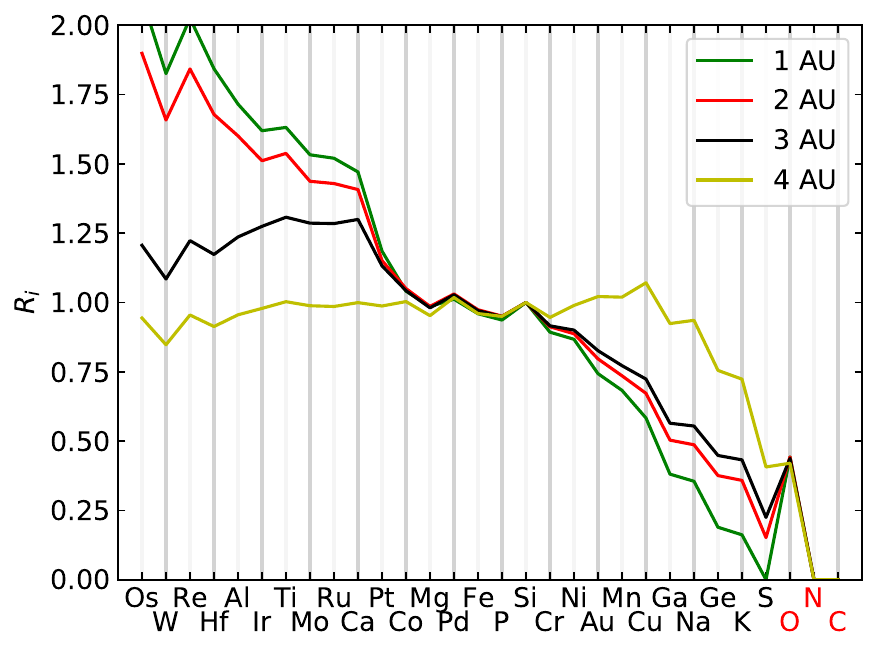}
	\includegraphics[width=0.34\textwidth]{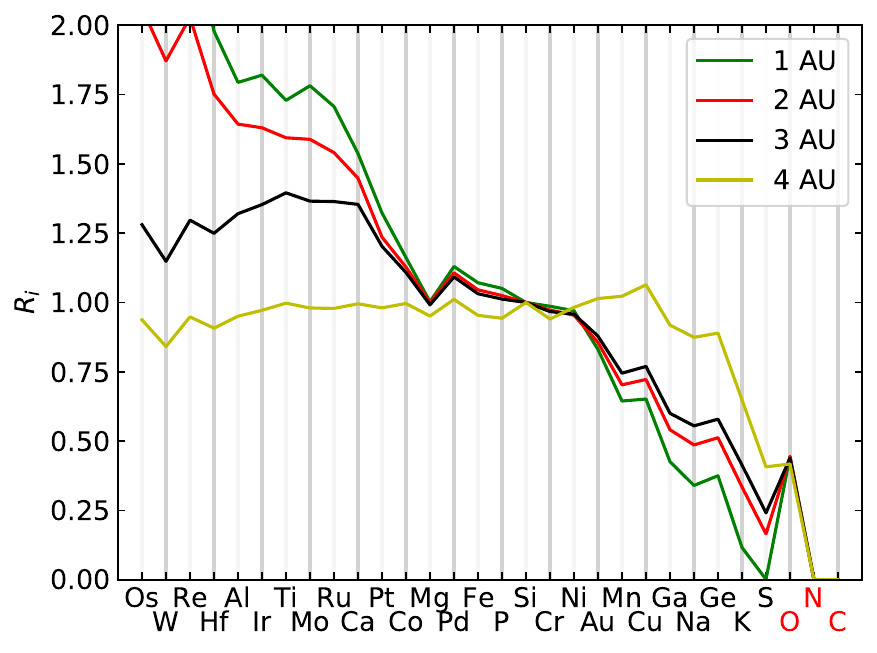}
	\includegraphics[width=0.34\textwidth]{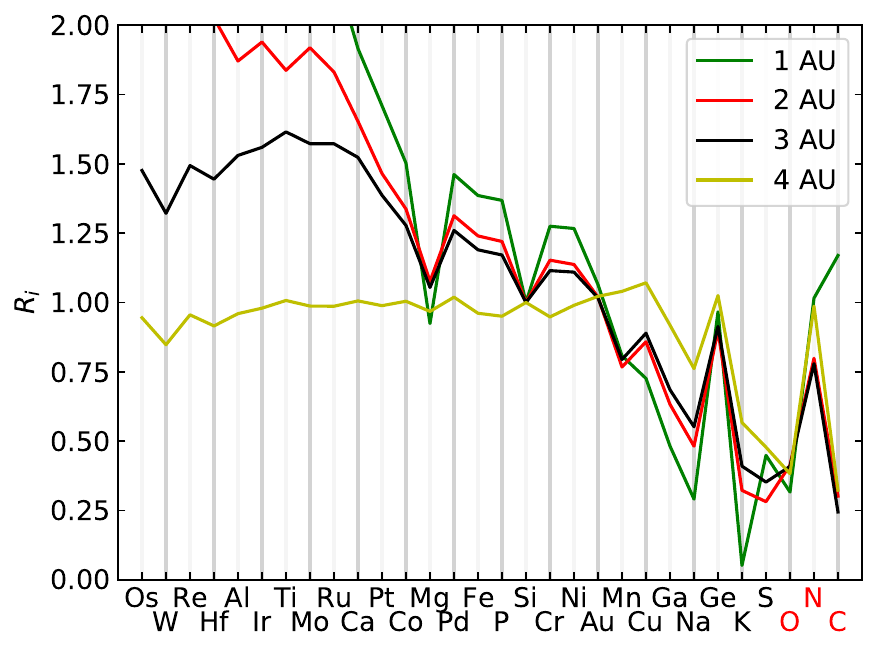}
	\includegraphics[width=0.34\textwidth]{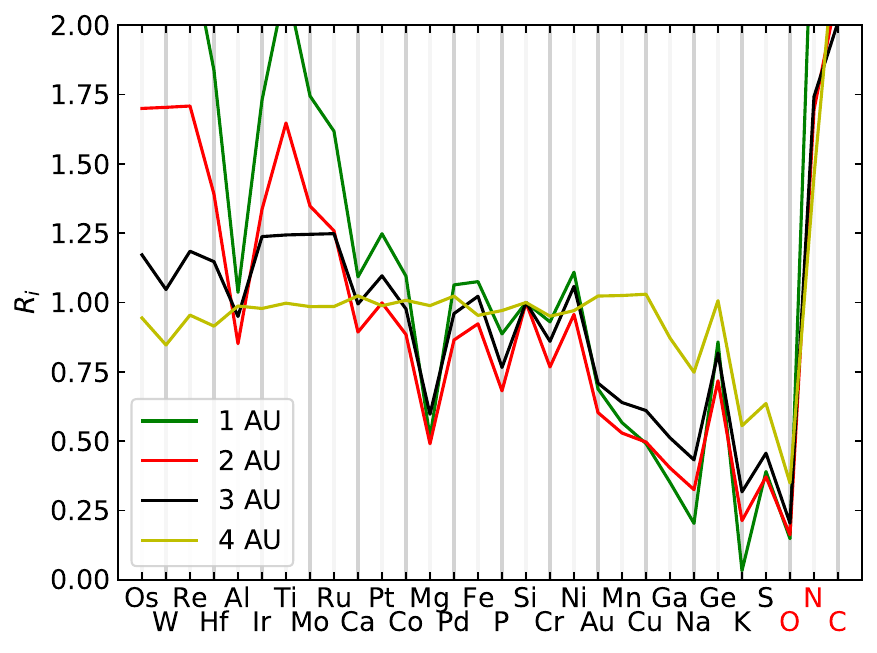}
    \caption{The same figure type as Figure \ref{fig:LiFig9CO} but examining N/O ratios. The top left, top right, bottom left, and bottom right correspond to the initial N/O ratios of 0.1, 0.5, 0.75, and 1.5, respectively. Note that the Solar N/O ratio is 0.138 and the Solar results are shown in the top left of Figure \ref{fig:LiFig9CO}.}
    \label{fig:LiFig9NO}
\end{figure*}
Figure \ref{fig:EleSig1} shows the final surface density of the five elements Si, C, O, Fe, and N. The plots show the same dichotomy on either side of the threshold C/O ratio as Figure \ref{fig:FracLine}. Figure \ref{fig:EleSig1} shows how the carbon condensation front (i.e. the areas where carbon condenses into solids) progresses to inner radii as the C/O ratio increases. The Solar simulation's elemental surface density on the left of Figure \ref{fig:EleSig1} is nearly the same as all simulations in the red-colored, low final C/O area of Figure \ref{fig:Frac3D}. Simulations with lower C/O ratios than the Sun do not exhibit the narrow region of condensed carbon around 6 AU, which can also be seen in Figure \ref{fig:Frac3D}. This narrow region of carbon is attributed to a form of iron carbide, Fe$_5$C$_2$. Also, note that the solid surface density of oxygen seen in the panels of Figure \ref{fig:EleSig1} is mostly attributed to silicates. See Section \ref{sec:Species} for a further breakdown of chemical species. We classify these similar systems as silicate-dominant and, in our simulations, the silicate-dominant behavior extends at least to a C/O ratio of 0.52. 

The middle of Figure \ref{fig:EleSig1} is the simulation with C/O=0.8 which is an intermediate system. These systems fall above a C/O ratio of 0.52 and end slightly above C/O=0.8. The characteristic feature of these systems is having a small but significant amount of carbon while silicate levels decrease slightly compared to the Solar model. Silicates remain dominant at inner radii as carbon approaches similar levels to oxygen at outer radii. The carbon here is mostly graphite but also contains carbides (for a full species breakdown, see Section \ref{sec:Species}). Silicate levels are slightly lower in the middle panel, which is best seen by comparing the oxygen profile to the iron profile. The iron profile remains nearly unchanged in the two left panels and serves as a good benchmark for comparison. The carbon condensation front does not yet reach the inner radii in intermediate systems. Thus, the reduction of silicate levels in the middle panel of Figure \ref{fig:EleSig1} causes lower surface densities at inner radii. This will be shown more clearly in Figure \ref{fig:SigmaCO} of Section \ref{sec:Other}.

The rightmost panel of Figure \ref{fig:EleSig1} is the C/O=1 simulation that is representative of the carbide-dominant systems, or the area of Figure \ref{fig:Frac3D} beyond an initial C/O of about 0.9. We classify these systems as carbide-dominant because graphite and carbides begin to exceed or match all other elements in the surface density profile. The increase in carbon coincides with an increase in silicon and nitrogen in some areas. Silicates fall to nearly zero at the inner radii while they remain a significant fraction at the outer radii. The rise in carbon more than compensates for the loss of oxygen at inner radii and high C/O ratios see a significant increase in solid surface density. \citet{Bond2010}'s `Sim.3' of HD19994, with a C/O of 1.26, shows a mixed composition planet that could be common at radii greater than 1 AU in our carbide-dominant models. The rightmost panel of Figure \ref{fig:EleSig1} shows that carbon and oxygen levels are similar beyond 1 AU. Our results suggest planets at outer radii will become progressively more carbon-rich at higher C/O ratios (seen previously in Figure \ref{fig:Frac3D}).

Above C/O=1, in Figure \ref{fig:EleSigMore}, decreases in final oxygen condensates become smaller -- significantly smaller than the corresponding decrease implied by the change in C/O ratio, in terms of molar surface density. Carbon-dominant models are unique in that excess carbon continues to condense as graphite. At C/O=1.5, Figure \ref{fig:EleSigMore} shows that carbon (as mostly graphite) reaches a molar density of 1 mol/cm$^2$, more than double any element from the oxygen dominant models. The increases in oxygen at very low C/O ratios, like C/O=0.1 and 0.01, are minor when compared to the Solar model. Oxygen-dominant models are fundamentally limited by the available reactants for oxygen to bind to and form condensates, usually as silicates. Increases in initial oxygen must coincide with increases in reactants for oxygen-dominant models to show significantly higher solid surface densities.

The middle and right panels in Figure \ref{fig:EleSig1} show that increasing the C/O ratio from the Solar value (0.501) to C/O=1 causes some nitrogen to condense into solid species, although still low compared to the total surface density. The same plot for the C/O=1 w/Oc simulation is nearly identical and the increased nitrogen is present in all high C/O simulations with ``constant sum'' and w/Oc. The increased nitrogen in these simulations is due to the formation of multiple metal nitrides, not nitrogen-oxygen or nitrogen-carbon species. Without the formation of these metal nitrides, this nitrogen would be lost as the disk evolves and the gas disperses. The rise of nitrides in high C/O ratios partly motivated our investigation into N/O ratios. The first N/O simulation with ``constant sum'' that has a significant change in elemental profile is N/O=0.75. However, all the simulations examining N/O ratios w/Oc have nearly identical dust compositions and profiles as the Solar simulation on the left of Figure \ref{fig:EleSig1}. This suggests that the N/O ratio with constant sum is most important in its indirect impact on the C/O ratio. Using the ``constant sum'' method, any increase in nitrogen comes with a loss of oxygen. Thus, the initial, indirect C/O ratio for N/O=0.75 with constant sum is about C/O=0.769.

In effect, since N/O w/Oc shows no differences, the N/O with ``constant sum'' simulations essentially behave as C/O with carbon constant (w/Cc) simulations since the amount of carbon is held constant at the Solar value. This will be further illustrated later in Section \ref{sec:Species}. In short, simulations that vary the N/O ratio show close agreement with the corresponding C/O simulation. In other words, an N/O simulation with an indirect C/O ratio is nearly identical to a C/O simulation of the same C/O ratio. We will continue to refer to these simulations as ``N/O with `constant sum''' but keep in mind that the impact appears to be solely due to carbon and the C/O ratio.

\subsubsection{Element Comparison over Ratios}
\label{sec:Ele3}

We use $R_i$, defined by Equation \ref{eq:Ri}, to compare many elements across many simulations and radii. This method was employed in \citet{Cassen1996} and used by \citet{Li2020}'s Fig. 9 to compare the Solar simulation results to chondrite compositions. We use a nearly identical figure to \citet{Li2020}'s Fig. 9 to make the comparison between different simulated ratios, and potential future studies, easier.

Figure \ref{fig:LiFig9CO} shows the $R_i$ for condensates of select elements at four locations in the disk. The elements are ordered left-to-right with decreasing $T_{50}$ temperatures. Oxygen, nitrogen, and carbon are appended to the right and highlighted in red. to indicate that these elements are being directly increased or decreased. However, to be clear, the $R_i$ is normalized to the simulation's local silicon abundance (see Equation \ref{eq:Ri}). Thus, a change in the shown $R_i$ can be caused by a change in the element's abundance and/or a change in the silicon abundance.

The top-left of Figure \ref{fig:LiFig9CO} is the same simulation in Fig. 9 of \citet{Li2020}, except here we do not include measured chondritic data points. Increasing carbon significantly changes the elemental compositions at the four distances indicated by the colored lines. However, certain elements are more impacted than others. Hf, Al, Ir, and Mg have large reductions with a C/O ratio of 1, shown in the bottom-left of Figure \ref{fig:LiFig9CO}. Ge is the only element with a large increase. Decreasing the C/O ratio to levels below the Solar value results in nearly identical plots to the Solar value.

Figure \ref{fig:LiFig9NO} shows the $R_i$ for condensates of select elements at four locations in the disk with varying N/O ratios. Many of the elements that are most impacted are the same from Figure \ref{fig:LiFig9CO}. The N/O ratio has little impact until the N/O ratio is 0.75, an increase by over a factor of five, in the bottom-left of Figure \ref{fig:LiFig9NO}. The change in behavior at an N/O of 0.75 with ``constant sum'' is caused by the indirect C/O ratio reaching $\sim$0.77, which is above the threshold value of 0.762 found previously in Figure \ref{fig:FracLine}. The top-right of Figure \ref{fig:LiFig9NO} has an N/O ratio of 0.5 with ``constant sum'', which means a C/O ratio of $\sim$0.66, but it does not differ significantly from the Solar $R_i$ results.

\subsection{Dominant Species}
\label{sec:Species}

The chemical species are often the true determinant of where in the disk the elements may condense as molecules in dust. Whether a chemical species is favorable depends on local conditions such as pressure, temperature, and elemental composition. The different elemental compositions of protoplanetary disks can enable
the condensation of particular molecular species that allow elements to condense where they normally would not with a solar composition. The increase in condensed nitrogen in the middle panel of Figure \ref{fig:EleSig1} due to metal nitrides is a good example of this important distinction. Combining volatile elements like nitrogen with refractories can increase the volatile composition from an elemental perspective, but the chemical species itself is not volatile. Chemical species that have high condensation temperatures and incorporate volatile elements can allow elements like nitrogen to be retained within minerals until a later time -- at which point chemical reactions may liberate it from its refractory partners. This is one example to illustrate the importance of dust condensation models that include robust chemistry models: robust in the thorough number of chemical species considered and robust in the quality of agreement with observations.

This section will consist of two parts, similar to the last two parts of Section \ref{sec:Elements}. The first, Section \ref{sec:Species1}, will cover where the most significant species condense with respect to radius. The second, Section \ref{sec:Species2}, will cover how the most significant species change with respect to the variation of initial elemental ratios.

\subsubsection{Species Distribution over Radius}
\label{sec:Species1}
Section \ref{sec:Elements} shows the increase in carbon's surface density. We now aim to examine the chemical species responsible in order to explain the elemental trends. Figure \ref{fig:Cspecs1} shows the most significant carbon-containing species that are affected by varying the initial elemental ratios. The three simulations shown are the same as Figure \ref{fig:EleSig1} and are in the same order. The left of Figure \ref{fig:Cspecs1} is the result of the Solar C/O ratio of 0.501 showing that carbon species just begin to condense near 6 AU at the end of our simulation as Fe$_5$C$_2$. The middle of Figure \ref{fig:Cspecs1} is the result of a C/O ratio of 0.8. The C/O=0.8 simulation is an intermediate system because the carbon condensation front has not reached its full inward extent, which is seen in the right panel where C/O=1. The inner region of the middle panel is still dominated by silicates (see Figure \ref{fig:EleSig1}) and the carbon levels are lower than the C/O=1 simulation. As with most carbide-containing models, the majority of the carbon condenses as graphite in C/O=0.8 and C/O=1. Graphite surface densities are about a factor of 10 larger than any other carbon species. Note that the solid surface density of a species may be higher than the total carbon surface density because of the non-carbon elements included in the mass of the species. 

Figure \ref{fig:Cspecs1} also helps to show the expansion of the carbon condensation front as the C/O ratio increases. When examining the condensation in the C/O ratios with a ``constant sum'' between C/O=0.52 and C/O=0.6, graphite begins to condense between 1.2-2.2 AU. Graphite expands to inner and outer radii with increasing C/O ratio, and the iron carbide spike seen in the left panel of Figure \ref{fig:Cspecs1} moves inwards. Between C/O=0.6 and C/O=1, the carbon condensation fronts of graphite and iron carbide meet, graphite becomes the dominant species until about 5 AU, and the inner edge of carbon consistently moves inwards while the oxygen condensation front consistently retreats to outer radii. Beyond C/O=1, graphite expands outward slightly to about 6 AU and iron carbide expands outward to about 8 AU. Inside of 6 AU, the graphite surface density consistently increases and the graphite surface density at 0.3 AU is about equal to that at 1 AU.

We use Figure \ref{fig:Nspecs} to examine which species are responsible for the large increase in nitrogen seen in Figure \ref{fig:EleSig1}. Figure \ref{fig:Nspecs} shows the most significant nitrogen-containing species for three simulations. The right two panels of Figure \ref{fig:Nspecs} are the C/O=0.8 and C/O=1 simulations and correspond in position to Figure \ref{fig:EleSig1} and Figure \ref{fig:Cspecs1}. The left panel is the simulation where N/O=1.5. We exclude the Solar simulation in the leftmost position of Figure \ref{fig:Nspecs} since condensed nitrogen-containing species are nearly non-existent and it is uninteresting. The nitrogen-containing species in the C/O=1 (right) and N/O=1.5 (left) simulations are nearly identical except for the lack of Si$_2$N$_2$O in the C/O=1 simulation. Si$_2$N$_2$O is a ceramic with dielectric properties. However, Si$_2$N$_2$O does appear with C/O ratios of 1.5 or greater. 

We conducted a C/O=1.1 w/Cc (with carbon constant to the Solar value) simulation specifically to determine if the increased nitrogen caused the increase in Si$_2$N$_2$O. The C/O=1.1 w/Cc simulation has a C/O ratio nearly equal to N/O=1.5's indirect C/O ratio of 1.10. The C/O=1.1 w/Cc simulation shows Si$_2$N$_2$O behavior almost identical to N/O=1.5. There are minor differences between the two, including slightly lower levels of Si$_2$N$_2$O in the C/O=1.1 w/Cc simulation. However, considering N/O w/Oc simulations show no difference to the Solar model, we attribute this minor difference to the small difference in initial oxygen abundance between N/O=1.5 and C/O=1.1 w/Cc. Thus, following this examination, we attribute the difference between C/O=1 with ``constant sum'' and N/O=1.5 to the differences in oxygen abundance. This is further evidence that increasing nitrogen has no impact on our models and that our N/O with ``constant sum'' simulations essentially function as ``C/O w/Cc'' simulations with the maximum amount of oxygen capped at $N(\text{O})=N(\text{O})_{\odot}+N(\text{N})_{\odot}$.

\subsubsection{Species Comparison Over Ratios}
\label{sec:Species2}
\begin{figure*}
	\includegraphics[width=0.96\textwidth]{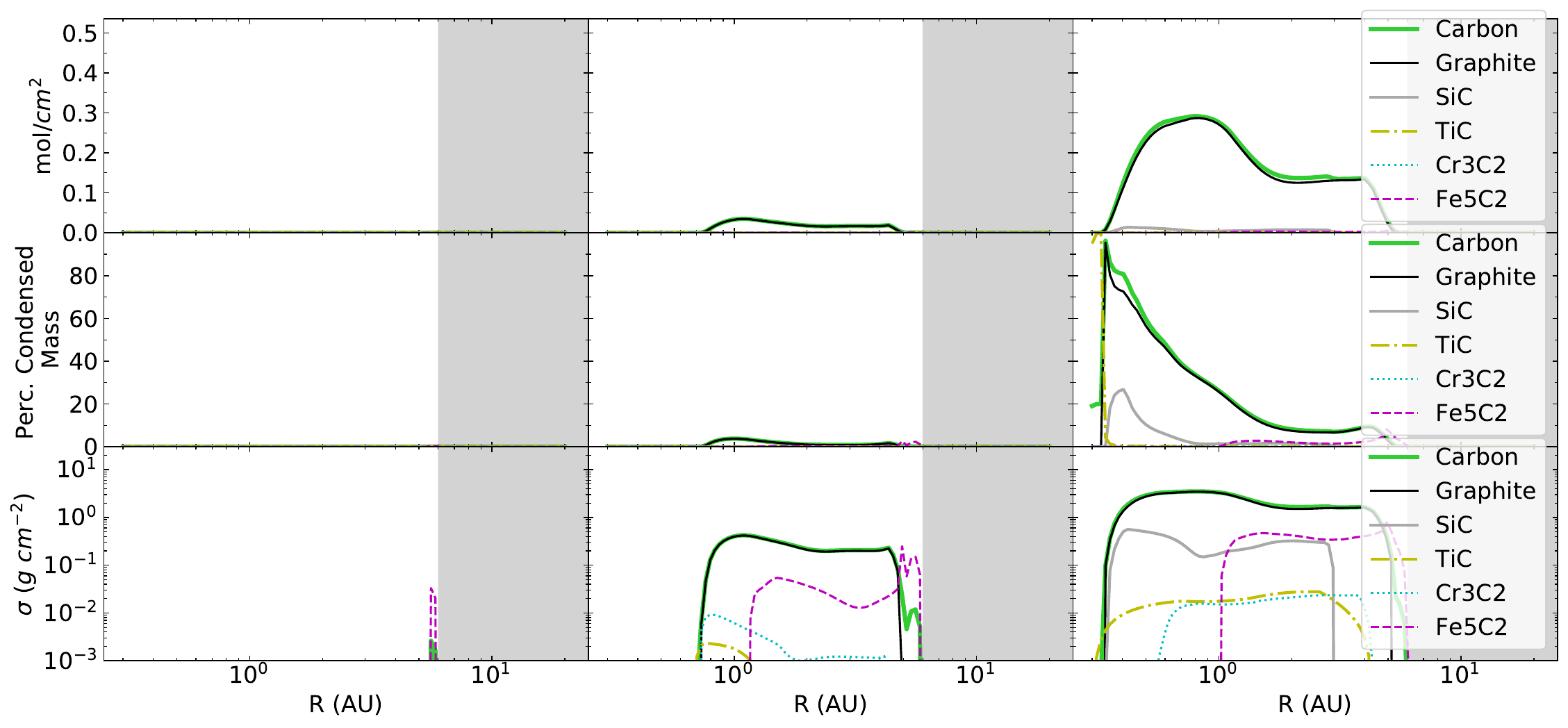}
    \caption{The solid surface density of carbon and the most significant species that comprise the total carbon in the three simulations shown in Figure \ref{fig:EleSig1}. (Left) Solar, (Middle) C/O=0.8, and (Right) C/O=1.}
    \label{fig:Cspecs1}
\end{figure*}

\begin{figure*}
    \includegraphics[width=0.96\textwidth]{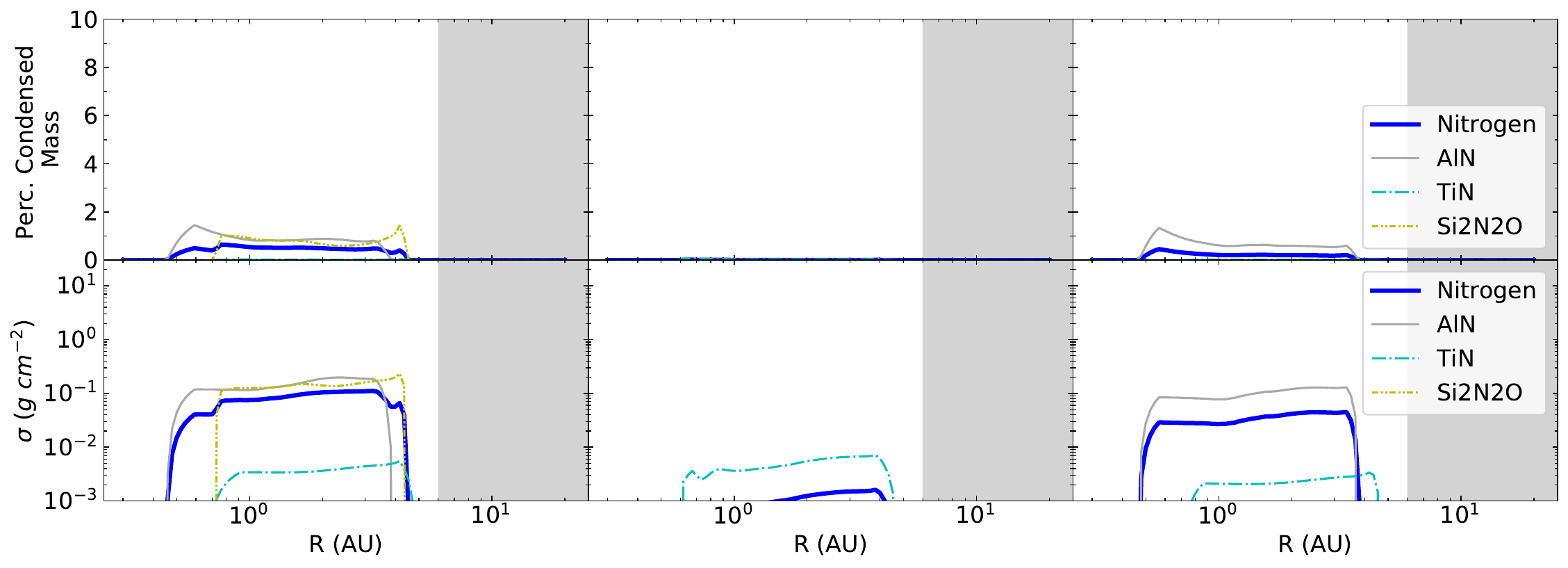}
    \caption{The nitrogen species that comprise the increased nitrogen in three simulations. Note the y-axis of the percent condensed mass plot is limited to 10\% here. (Left) The N/O=1.5 simulation. Note that N/O=1.5 is positioned on the left to avoid being misidentified as C/O=0.8 or C/O=1, which hold the middle and right positions, and because the Solar plot is not shown here. (Middle) The C/O=0.8 simulation (the middle panel of Figure \ref{fig:EleSig1}). (Right) The C/O=1 simulation, shown here, is the C/O simulation that has the closest C/O ratio to N/O=1.5's indirect C/O ratio. N/O=1.5 has an indirect C/O ratio of 1.10.}
    \label{fig:Nspecs}
\end{figure*}
\begin{figure*}
	\centering
	\includegraphics[width=0.96\textwidth]{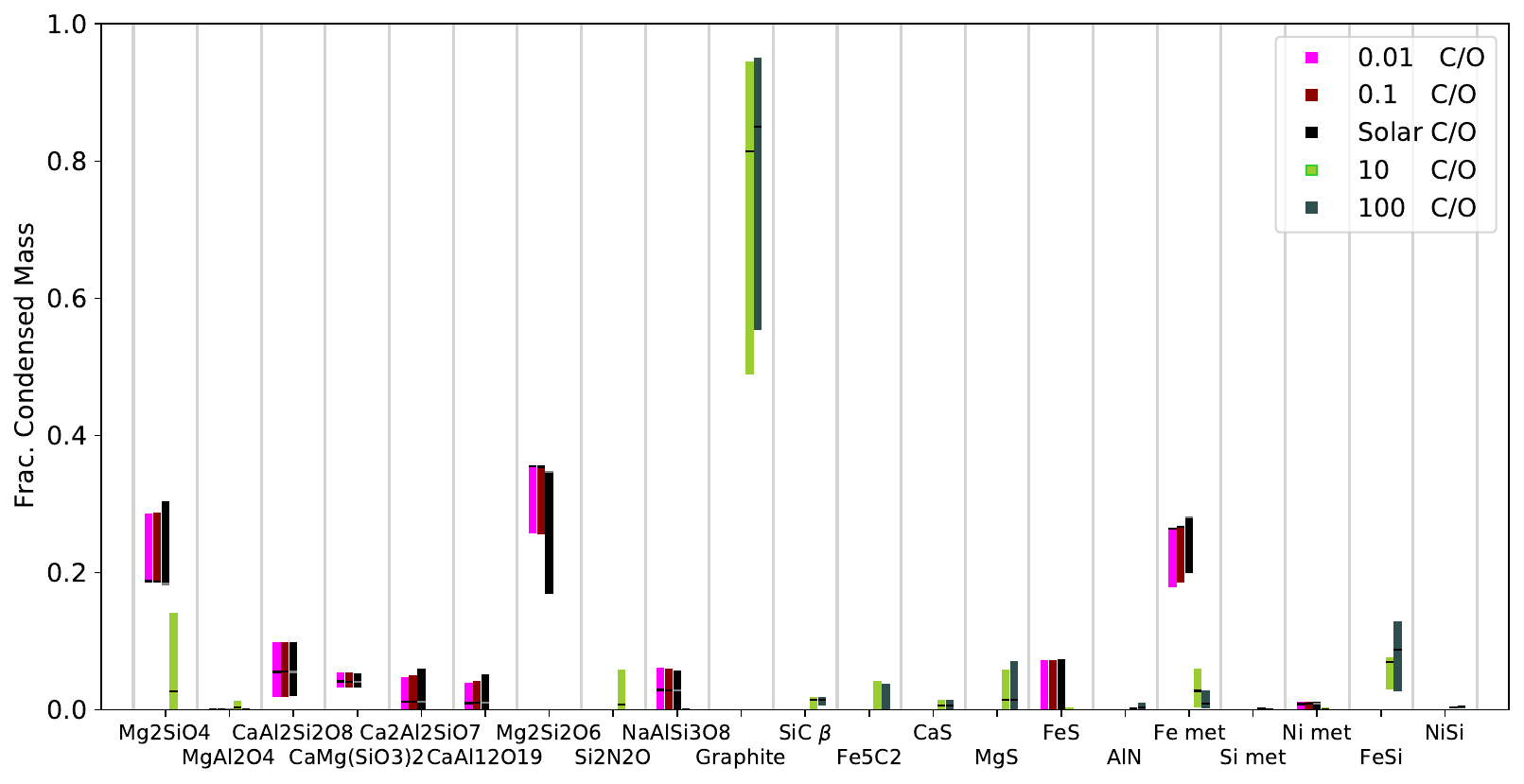}
    \caption{The fraction of condensed mass comprised of major species at radii between 0.5 AU and 4 AU. Vertical bars span between the maximum and minimum of mass fraction for the species measured at 0.5, 1, 2, and 4 AU for the respective simulation. Each vertical bar is offset from the center of the species name in the same repeating order (left-to-right) as the order of the legend (top-to-bottom). Black tick marks (grey ticks for the black/Solar bar) denote the value at 1 AU. For example, the bottom of the C/O=10 line above the species graphite indicates the minimum condensed mass fraction of graphite at radii of 0.5, 1, 2, and 4 AU was about 50\%. The top of the line for C/O=10 shows the maximum was above 90\%. Note that only the values associated with the 1 AU tick marks can be summed to find the total condensed mass of the shown species.}
    \label{fig:SpecCnameEXTR}
\end{figure*}
\begin{center}
\begin{figure*}
    \centering
	\includegraphics[width=0.96\textwidth]{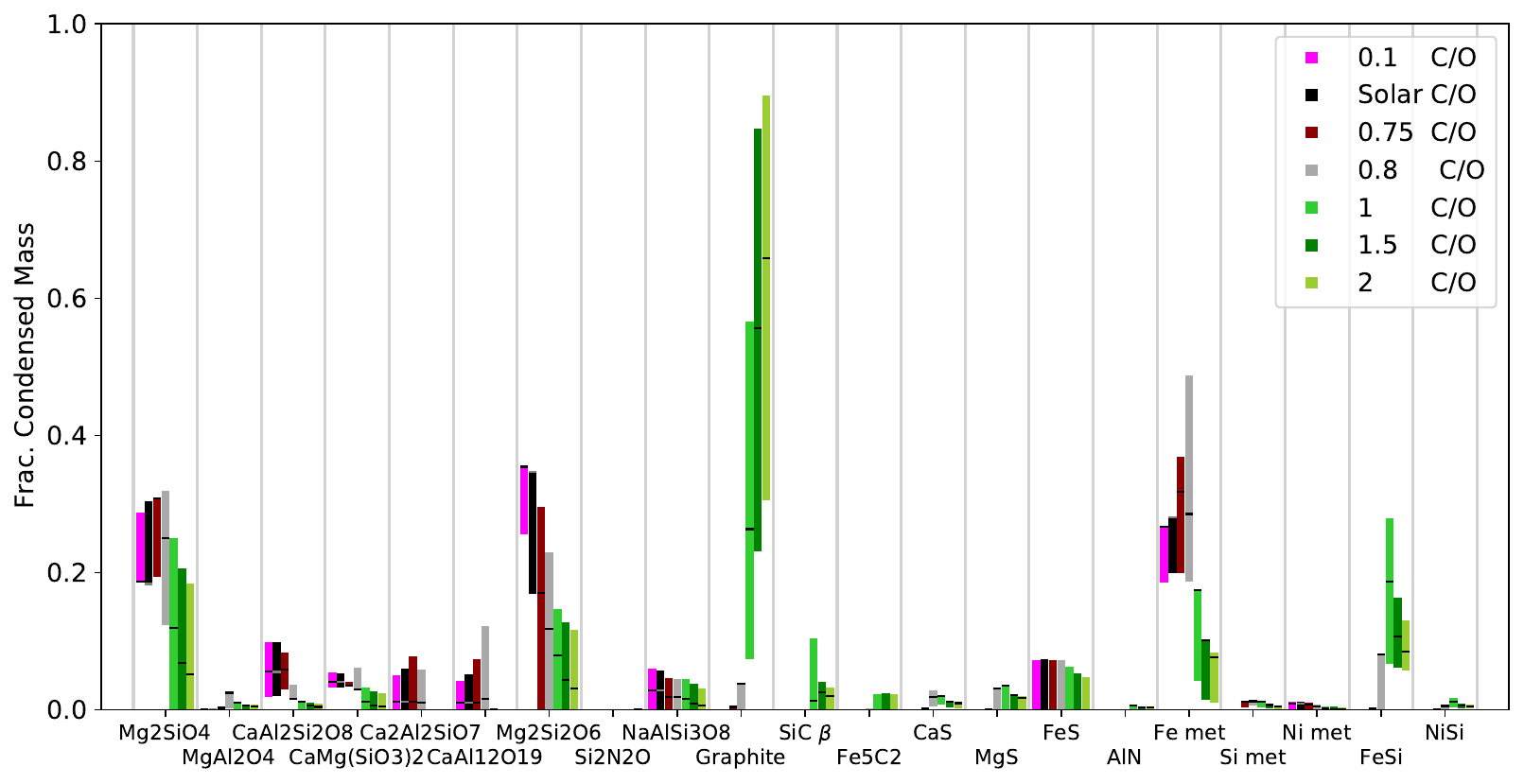}
    \caption{The same graphic as Figure \ref{fig:SpecCnameEXTR} but here the C/O ratios examine the transition in composition around C/O=0.8.}
    \label{fig:SpecCname}
\end{figure*}
\end{center}
\begin{figure*}
	\includegraphics[width=0.96\textwidth]{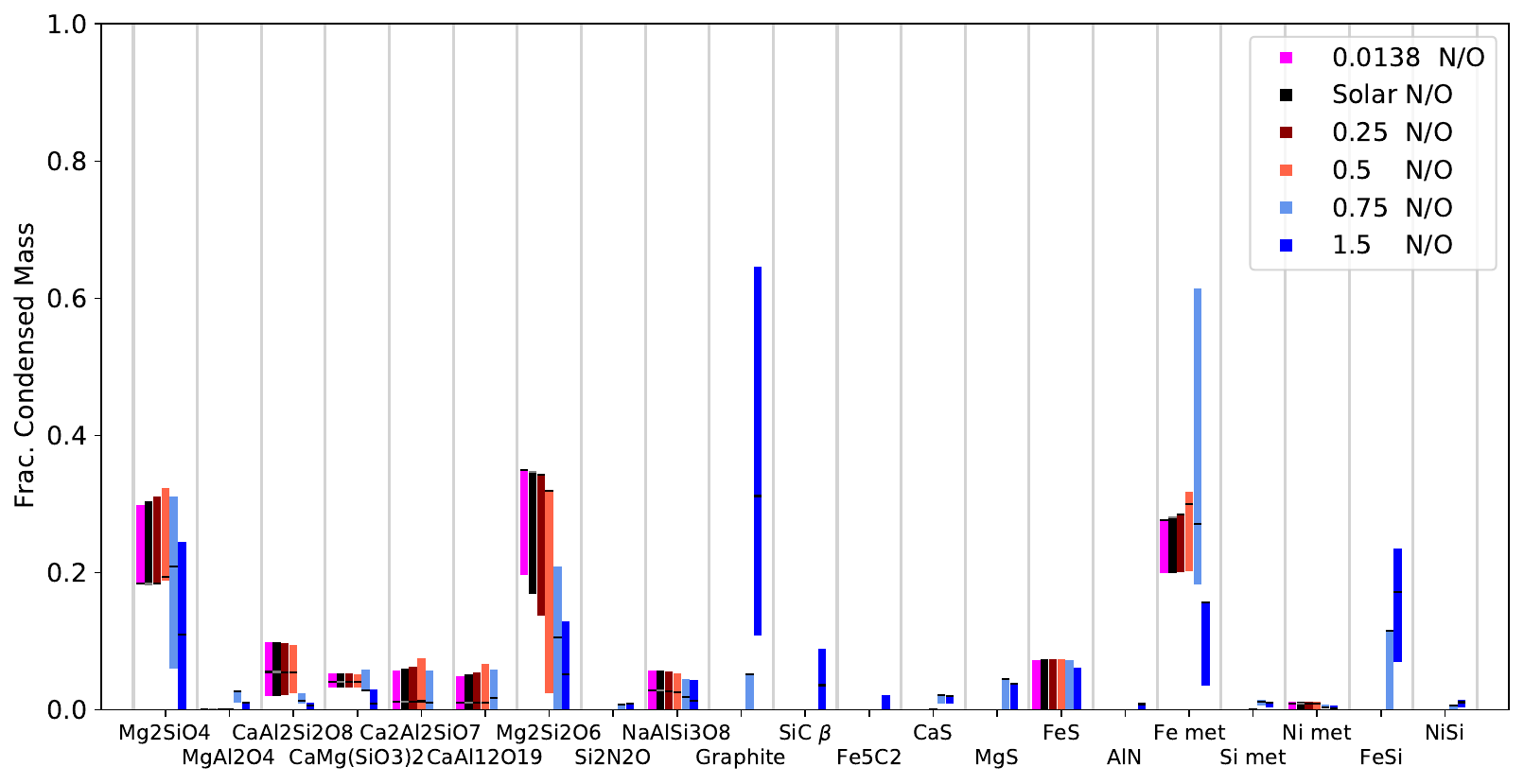}
    \caption{The same figure as Figure \ref{fig:SpecCname} except N/O ratios are varied for each simulation. Vertical bars span between the maximum and minimum mass fraction for the species measured at 0.5, 1, 2, and 4 AU for the respective simulation. Black tick marks (grey ticks for the black/Solar bar) denote the value at 1 AU.}
    \label{fig:SpecNname}
\end{figure*}

We attempt to provide a comprehensive breakdown of the species impact by using Figures \ref{fig:SpecCnameEXTR}-\ref{fig:SpecNname}. Figure \ref{fig:SpecCnameEXTR} helps to identify some distinctions between systems with high or low C/O ratios. Here we examine a set of C/O ratios that span into extreme values with a Solar C/O ratio for reference. This figure demonstrates the convergent end behavior of varying C/O ratios around a Sun-like star. The condensed material around a star with a Solar C/O ratio closely aligns with the silicate-dominant models. All of the models shown in Figure \ref{fig:SpecCnameEXTR} have condensed material dominated by either two or three species. The carbide-dominant models are dominated by graphite and FeSi (iron silicide). The silicate-dominant models are dominated by Mg$_2$SiO$_4$ (fosterite), Mg$_2$Si$_2$O$_6$ (clinoenstatite), and Fe metal. Si$_2$N$_2$O is only a large fraction at a C/O ratio of 10 and appears in Figure \ref{fig:SpecNname} at an N/O ratio of 0.75 and 1.5. Si$_2$N$_2$O is present in high C/O with ``constant sum'' simulations, as mentioned in the previous section and seen in Figure \ref{fig:Nspecs}. However, Si$_2$N$_2$O and other species compose a small fraction of total mass because of the high amounts of graphite present in high C/O simulations.

Silicate-dominant models, including the Solar C/O ratio, show a more varied composition than the carbide-dominant models. Species containing iron show a particularly dichotomous feature. In silicate-dominant models, iron is found in Fe metal and FeS. High C/O models have iron in Fe metal, FeSi, and Fe$_5$C$_2$ with significantly more magnesium content in MgS. With a C/O ratio of 10, most remaining oxygen appears to be contained in Mg$_2$SiO$_4$. The increased MgS abundance in the carbide-dominant models is explained by the decrease in Mg$_2$SiO$_4$, Mg$_2$Si$_2$O$_6$, and CaMg(SiO$_3$)$_2$. Mg$_2$SiO$_4$ and Mg$_2$Si$_2$O$_6$ fractions decrease at high C/O ratios but are still present in significant amounts. SiC is interesting in that it only composes large fractions at a C/O ratio of around 1.

The absolute amount of SiC consistently increases with the C/O ratio but the increase in graphite surface density reduces the relative abundance. Similar to SiC, other low-abundance species in the carbide-dominant regimes have much higher abundances relative to Solar models. Like Si$_2$N$_2$O in the previous paragraph, many species appear so sparse because graphite is so prevalent in comparison. Note that this is one shortcoming of Figure \ref{fig:SpecCnameEXTR}-\ref{fig:SpecNname}'s method of presentation; graphite levels become so high in certain areas and simulations that the silicates will become a small fraction of the total mass even if silicates remain at surface density levels similar to the Solar model. This is particularly an issue for very high C/O ratios, like 10 and 100, but also occurs at intermediate C/O ratios where graphite reaches levels comparable to silicates and silicates are still prevalent (e.g the middle panels of Figures \ref{fig:EleSig1} and \ref{fig:Cspecs1}). This issue illustrates the importance of our examination of the impacts of the C/O ratio using the fraction of mass but also the surface density.

More probable stellar C/O ratios may range from 0.1-2. Figure \ref{fig:SpecCname} spans this range showing that the condensed material is still largely confined to the same five species from Figure \ref{fig:SpecCnameEXTR}, although more evenly distributed. These models span the area around the threshold C/O ratio where the chemical makeup is a mix between the carbide-dominant and silicate-dominant regimes. C/O ratios between 0.75-1 are particularly sensitive as the graphite fraction at 1 AU goes from $\sim0\%$ to $>20\%$. FeS is common in all simulations around the threshold C/O ratios shown in Figure \ref{fig:SpecCname}.

\begin{figure*}
    \centering
    \includegraphics[width=0.96\textwidth]{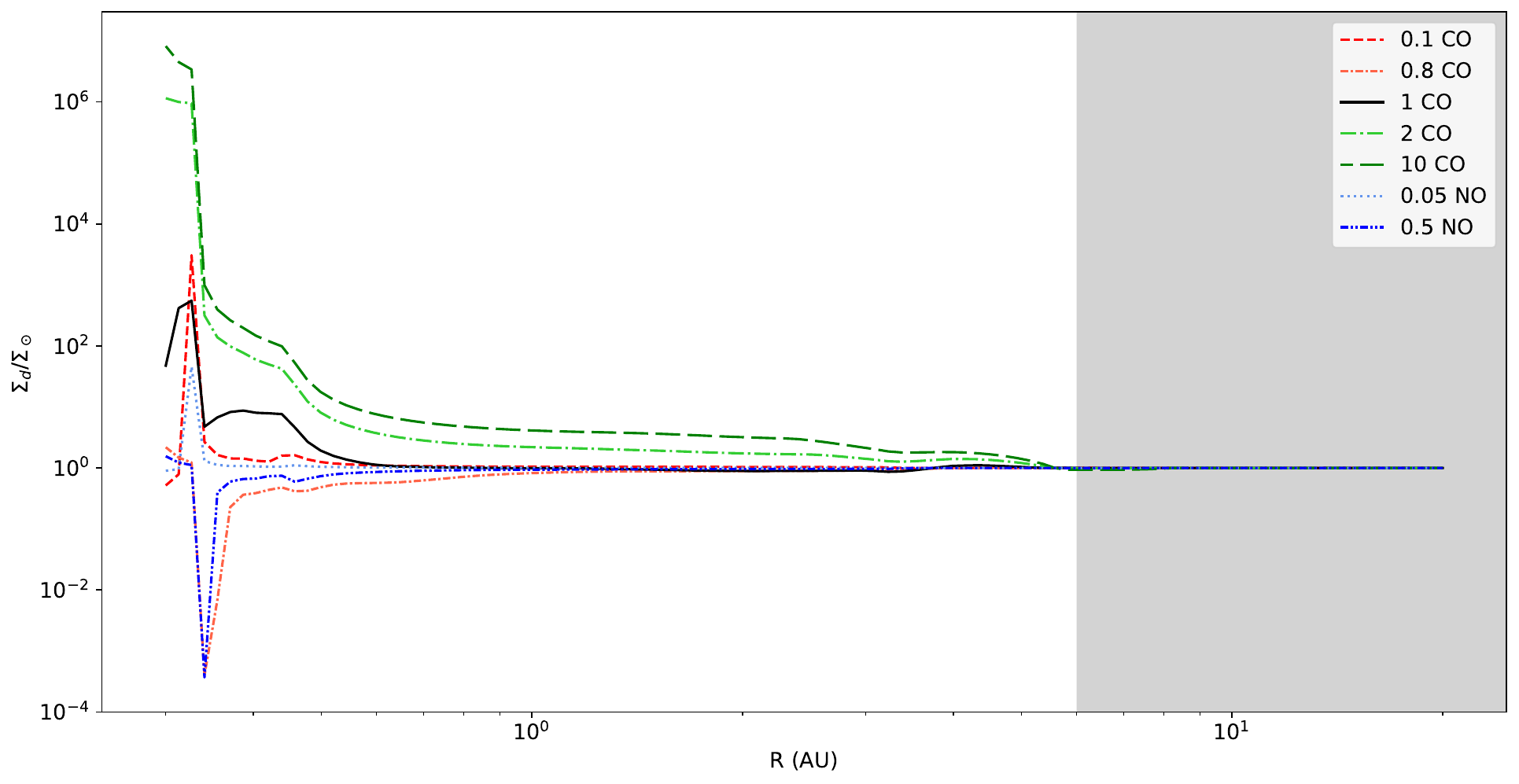}
    \caption{The solid surface density profile with distance from the Sun. The solid surface density for each C/O ratio simulated is normalized to the solid surface density profile of the Solar C/O ratio.}
    \label{fig:SigmaCO}
\end{figure*}

Figure \ref{fig:SpecNname} is identical to Figure \ref{fig:SpecCnameEXTR} except that we examine a range of N/O ratios with ``constant sum''. High N/O ratios exhibit significant amounts of graphite, SiC $\beta$, and FeSi. Fe metal also remains a significant species with a large increase at an N/O ratio of 0.75. This large relative increase in Fe metal is located around 2 and 4 AU. The increase of Fe metal for an N/O ratio of 0.75 is, in part, caused by a decrease in the local solid surface density of other species. Some oxygen-containing species still remain in significant amounts at an N/O ratio of 1.5. Mg$_2$SiO$_4$ and Mg$_2$Si$_2$O$_6$ follow a similar behavior here as the carbide-dominant models. The N/O w/Oc simulations show no noticeable change across all of the N/O w/Oc ratios examined. The simulation results for N/O w/Oc are excluded from our figures as they appear identical to the Solar model for all figures. The results of the two N/O varying methods suggest that the main impact of varying the N/O ratio with ``constant sum'' is the decrease in oxygen and, thus, the indirect increase in the C/O ratio.

\subsection{Other Impacts}
\label{sec:Other}
One consequence of increasing the C/O ratio is a general increase in the final solid surface density. Figure \ref{fig:SigmaCO} shows the total solid surface density normalized to the Solar model. High carbon and high oxygen levels both cause increased surface density at some locations. However, high carbon levels cause a significantly larger increase in surface density near the star that approaches lower values, comparable to the Solar profile, only around 6 AU. The high surface densities for C/O=2 are almost entirely associated with graphite. For C/O=1, as shown in Figure \ref{fig:Cspecs1}, the high inner surface density is due to carbides while the outer surface density is a combination of silicates and carbides.

Note that a relative surface density of $10^7$ for C/O=2 and C/O=10 is, in part, a consequence of the Solar solid surface density being so low at small radii in our models. The large increase of inner surface density because of high C/O ratios is also seen in \citet{Bond2010}. In Fig. 15 of \citet{Bond2010}, their HD4203 simulation (with a cited C/O=1.86) is very similar to our C/O=2 simulation in Figure \ref{fig:SigmaCO}. Our results differ on C/O=0.8, where \citet{Bond2010}'s HD177830 (C/O=0.83) shows increased surface density while C/O=0.8 in Figure \ref{fig:SigmaCO} shows a decreased surface density compared to the Solar simulation. The decrease our model sees when C/O=0.8 is due to a decrease in silicates at inner radii while carbide levels have not yet counteracted this loss of solids. When C/O=1, carbides have reached high enough levels to both replace the mass lost from the decrease in silicates and surpass the Solar surface densities. \citet{Bond2010} attributes the increased surface density of HD177830 to the increase in the Mg/Si ratio. \citet{Bond2010}'s HD4203 (C/O=1.86) has a nearly Solar Mg/Si ratio, thus, we can be confident that the primary cause of its increased surface density is due to the increased C/O ratio.

\section{Discussion}

\subsection{Trends in C/O Ratio Behavior}

The change in behavior from silicate-dominant to intermediate systems begins between C/O=0.52 and 0.6, and intermediates end at about 0.9. The historically cited threshold value of C/O=0.8 aligns with our condensation simulations where 1-10\% of the condensed disk mass is composed of carbon. These carbon levels are much higher than what we observe in the inner Solar system, but the vast majority of the material has nearly the same composition as the silicate-dominant systems, seen in Figure \ref{fig:SpecCnameEXTR}. The more pressing impact of C/O ratios around the threshold value of 0.8 may be the significant decrease in total surface density shown in Figure \ref{fig:EleSig1} and \ref{fig:SigmaCO}. The threshold value of $\sim$0.8 is a good indicator for when changes in condensation behavior begin, but C/O values greater than 0.9 -- where intermediate systems end --  may be a better indicator for when systems are assuredly carbide-dominant. In other words, making a confident prediction that a stellar system is carbide-dominant should rely on a C/O ratio greater than 0.9. Otherwise, for C/O ratios above $\sim$0.6 but less than 0.9, condensation models would be needed for a confident prediction of the stellar system's likely composition.

Increasing the C/O ratio beyond C/O=1 continues to increase graphite but the decrease in silicates becomes smaller than the proportional change in oxygen according to the intial C/O ratio -- from a molar perspective in Figures \ref{fig:EleSig1} and \ref{fig:EleSigMore}. The molar surface density of oxygen in the Solar model is about 0.3 mol/cm$^2$ at 1 AU and falls to about 0.1 mol/cm$^2$ for the C/O=1 model. At C/O=1.5, oxygen remains at a surface density of about 0.1 mol/cm$^2$ while graphite reaches 1 mol/cm$^2$. The underlying level of silicates seems to reach a sort of minimum at C/O ratios above 1. Some areas of the disk have nearly as much silicates as the Solar model, but these silicates are mixed with equal or larger amounts of graphite.

Decreasing the C/O ratio does not cause large increases in surface density. This is likely because the major carrier of condensed oxygen, silicates, relies on other elements. This may suggest that other elemental ratios can be important for oxygen-dominant and intermediate systems. However, carbon-dominant systems may be less sensitive to other elemental ratios because graphite only relies on carbon. This is another reason to be wary of classifying the composition of stellar systems without the use of condensation models when the C/O ratio is above $\sim$0.6 but less than 0.9.

\subsection{Implications for Abundant Graphite}

The concept of carbon planets was first discussed by \citet{Kuchner2005} after being inspired by \citet{Lodders2004}'s suggestion that Jupiter formed from more tar than ice. The concept of a tar line, composed of hydrocarbons and organics, has become prevalent. This tar would be present below temperatures of 450 or 350 K and some of these organics could be preexisting material that condensed long before the Solar system \citep{Lodders2004}. Our model does not include these tars and graphite composes most of the condensed carbon in our simulations. Although, it is safe to assume that some of the graphite in the colder regions of our model would be replaced or supplemented by tars.

Both our results and \citet{Bond2010}'s show high C/O ratios can significantly increase the disk's solid surface density. High solid surface density regions would likely form planetesimals more efficiently. This could lead to larger planets or more planets. \citet{Bond2010} suggested the formation of carbide planets. Our simulations show areas that would form majority-carbide planets (i.e. with very little silicates) would have significantly higher surface densities than the Solar system. This may imply that majority-carbide planets might be more massive and more likely to form as super-Earths or larger. Jupiter and the other giant planets in our Solar system are enriched in carbon. If \citet{Lodders2004} is correct that carbonaceous material played a significant role in Jupiter’s formation, then stellar systems enriched in carbon could host more giant planets.

Regions that have surface densities similar to the Solar model, outside of about 0.5 AU in Figure \ref{fig:SigmaCO}, often have a mixed composition in our models. Figure \ref{fig:EleSig1} shows that oxygen, and hence silicates, become significant outside of 0.7 AU. Planets in carbide-dominant systems that lie in the habitable zone of a Solar-like star (i.e. approximately between the position of Venus and Mars) would be composed of mixed planets that have significant levels of silicate in addition to a majority graphite composition. Mixed planets would have different chemistry, and likely different geology, when compared to either silicate planets, like Earth, or the carbide planet structure that \citet{Bond2010} suggests. There may be a rich spectrum of planet chemistry and geology depending on the final carbon mass fraction of a planet. The Solar system's terrestrial planets are all silicate-dominant and the remaining, relatively small, differences result in the variety we observe between them. Our view of common planet structure, chemistry, and composition may be quite limited by the lack of planets containing significant (greater than $\sim$1\%) amounts of carbon in our Solar system.

\subsection{Impact of the C/O Ratio on Other Elements}

Nitrogen-containing condensates are more abundant at high C/O ratios. Nitrogen can make up to 0.5\% of the local disk mass when C/O=1 in Figure \ref{fig:Nspecs}. Although nitrogen abundances remain orders of magnitudes smaller than major elements, 0.5\% nitrogen is far higher than the Earth's abundance. Furthermore, this increased nitrogen is delivered in a different form than what is commonly seen on Earth. While it is difficult to directly compare with \citet{Mori2014}'s Fig. 1, their high C/O models also show an increase in nitrogen.

Other trace elements can be impacted significantly, as discussed in Section \ref{sec:Ele3} and shown in Figure \ref{fig:LiFig9CO}. However, our model does not include dissolved gases in solid material nor does it include an extensive amount of hydrated minerals. Graphite and SiC, the prevalent species of the inner regions of carbide-dominant systems, will dissolve different gases at different concentrations, incorporate different hydrated minerals, and facilitate different reactions after dust condensation has finished.

\subsection{Changes after Condensation}

The specific species we find in our dust condensation models may change significantly due to chemical reactions after condensation. Chrondrite formation and hydration with water vapor could facilitate some changes. Planetesimal accretion can create high enough heat and pressures to change chemical species and cause volatile depletion. Once planetesimals accrete enough mass for differentiation, further changes in the exact species can occur in the planet's interior.

\citet{Allen2020} conduct high-pressure experiments to examine potential chemical reactions that can occur on carbide-dominant planets. They assume the planets are initially composed of a graphite rind crust, a SiC mantle, and an Fe-C alloy core. If water or hydrated minerals are then delivered to the planet in a late veneer, chemical reactions can occur at depth to produce a silicate crust and an upper mantle composed of diamond and dense hydrous silica. \citet{Allen2020} also find that these reactions would produce methane and H$_2$ gas, depending on the depth of the reactions. Our models suggest that these planets may often have more graphite than \citet{Allen2020} expects. However, their study highlights the different chemistry that may occur within these planets and further affect the chemical profile in unexpected ways.

All of the previous examples should have little impact on the bulk elemental composition of the planets. Planetesimal migration or scattering can move planetesimals away from their initial formation location and result in compositions inconsistent with a planet's final location. Early giant planet formation and disk dynamics would likely determine whether the sharp radial changes in the C/O ratio seen are reflected in the final planet compositions. There are also commonly discussed examples of volatile depletion through out-gassing and photoevaporation. Fragmentation of differentiated bodies may provide another route for element loss in some bodies through collisional erosion (e.g. \citet{Campbell2012},\citet{Allibert2021}). However, impacts that significantly change the bulk composition through loss of material could be rare. Impacts and fragmentation are generally known to mix material from different locations in the disk (e.g. \citet{Childs2023}).

\subsection{Differences in Models}

Our dust condensation model uses a more comprehensive chemical condensation code, GRAINS \citep{Petaev2009}, and a different disk evolution model than both \citet{Bond2010} and \citet{Mori2014}. While \citet{Bond2010} and \citet{Mori2014} use the same chemical condensation software, HSC Chemistry, their disk evolution model and planet formation models are different. These differences in the underlying models are expected to produce differences in results. However, all of the models consistently demonstrate certain qualitative results: moderately higher C/O ratios than the Sun's produce large differences in terrestrial planet composition, and the composition of the planets in these high C/O systems is more sensitive to orbital radius than in low C/O systems.

Our high levels of carbon are consistent with \citet{Bond2010}'s results in the bottom panel of their Fig. 9. The rightmost panel of Figure \ref{fig:EleSig1} also shows iron and silicon mass fractions that are agreeable with \citet{Bond2010}'s Fig. 14. However, \citet{Bond2010}'s models do not show similar levels of TiC, AlN, or TiN in their model results despite including these species in their chemistry model. Quantitative comparisons are difficult because of the significant differences in models, but especially due to different methods of presentation. \citet{Mori2014}, \citet{Bond2010a}, and \citet{Bond2010} use Solar abundances that put the Solar C/O ratio at about 0.54 which is higher than ours at 0.501 via \citet{Lodders2003}. The disk evolution models and the dust condensation models that each study uses may be equally plausible but they are indeed different and can cause different results. \citet{Mori2014}'s Fig. 1 shows molar condensate composition over radius, which is straightforward other than its arbitrary scaling. However, \citet{Mori2014}'s Fig. 2 actually shows the planetesimal composition which may be additionally impacted by \citet{Mori2014}'s parameter of planetesimal formation rate. \citet{Bond2010} and \citet{Jorge2022} show the condensate composition but only as a function of temperature. Furthermore, the above-mentioned studies do not use the same scaling in their presentation of composition results. Thus, it is difficult to make a direct comparison between their surface density profiles and ours.

\citet{Mori2014}'s presentation of results is most easily comparable to ours. Our results show silicate-dominant behavior ending between C/O=0.52 and 0.6 with the condensation of small amounts of carbon. \citet{Mori2014} shows that carbon begins composing a few percent of condensed mass at certain radii at C/O=0.54 and 0.6. \citet{Mori2014}'s results also show a similar increase in condensed nitrogen with high C/O ratios. The C/O ratio of these behavior changes agrees quite well, however, \citet{Mori2014} finds their carbon first condensing at inner radii while we find it first condensing at outer radii. This discrepancy may be due, in part, to \citet{Mori2014}'s Fig. 2 being affected by their parameterized planetesimal formation rate. However, \citet{Mori2014}'s Fig. 1 also shows that carbon condensation as dust occurs more predominantly at inner radii and mostly within 3 AU.

\citet{Mori2014}'s Fig. 2 shows that their model predicts C/O=0.8 produces planetesimals with carbon weight percents greater than 20\% inside of 0.3 AU, which rises to about 50\% at 0.1 AU. This may be impacted by their planetesimal formation rate, but our C/O=0.8 model predicts carbon will not surpass 10\% weight percent at any point in the disk. When the C/O ratio increases to C/O=1, our model predicts more carbon than \citet{Mori2014}. Our C/O=1 results show that carbon, mostly in the form of graphite, can reach 80\% of the disk mass around 0.4 AU and remains above 10\% of disk mass out to 4 AU (see Figure \ref{fig:EleSig1}). \citet{Mori2014}'s Fig. 2 shows their C/O=1 model peaks at about 60\% carbon at 0.1 AU, falls below 10\% by 0.4 AU, and reaches $\sim$0\% slightly outside of 2 AU.

The implications of these differences mean that \citet{Mori2014} predicts carbide-dominant planets will form at lower C/O ratios than we predict, but these carbide-dominant planets, above $\sim$30\% carbon, will form exclusively within 0.3 AU. We predict planets above $\sim$30\% carbon will form within about 1 AU at C/O=1, but intermediate system planets around C/O=0.8 will have carbon weight percents lower than 10\%. Additionally, our C/O=0.8 model shows the most carbon weight percent between about 0.8 AU and 4 AU. In summary, our model may suggest that more significant changes occur between C/O=0.8 and C/O=1 while \citet{Mori2014} find that carbon is about as significant at C/O=0.8 as it is when C/O=1.

For Solar-like stars, with habitable zones centered around 1 AU, our model predicts an overlap between the areas where carbide-dominant planets form and where liquid water may exist. Our models increase the potential impact of carbide-dominant planets on habitability as previous studies do not show as much overlap with the habitable zone. Note that our results cannot be directly applied to stars with significantly different masses than the Sun. Low-mass stars, for example, may have different surface density and pressure-temperature profiles. Our dust condensation model was combined with a pebble accretion model to study planet formation in the M-dwarf, TRAPPIST-1 system \citep{Childs2023}. However, we have not yet implemented a general model for such low-mass star systems.

\section{Conclusions}

We model sequential dust condensation in an evolving protoplanetary disk to examine the impact of the C/O ratio on the composition of solids around a Solar-like star. We describe three different system types in our findings. The Solar system falls into the silicate-dominant, low C/O ratio systems. Silicate-dominant systems end at a stellar C/O ratio somewhere between 0.52 and 0.6. At C/O ratios between about 0.6 and 0.9, we have intermediate systems. Intermediate systems show a decrease in silicates while carbides begin to become significant -- primarily at inner radii where silicon begins binding predominantly with carbon and other metals. Carbide-dominant systems begin around a C/O ratio of 0.9. Carbide-dominant systems are defined by high carbide surface densities at inner radii and carbide levels comparable to silicate levels at outer radii. As the C/O ratio increases further, carbide-dominant systems show large increases in graphite while other elements like iron, oxygen, and silicon remain largely the same between C/O=1 and C/O=2. Systems above a C/O ratio of 2 are rare if current observations are reliable \citep{Bond2010}.

The threshold C/O ratio, historically cited as around 0.8, indicates when condensate composition begins to change and carbon reaches about 1-10\% of total disk mass in our models. However, our models show more significant changes occur between 0.8 and 1 where graphite becomes a major component. This finding is in contrast to previous studies, like \citet{Mori2014}, which find that carbon is about as significant at C/O=0.8 as it is when C/O=1. Beyond C/O=1, we find graphite continues to increase but other element and species surface densities remain largely constant.

Graphite surface densities increase to high levels at inner radii when the C/O ratio reaches C/O=1 or more. When the C/O ratio exceeds C/O=1, graphite becomes the most significant species at most locations in the disk. This graphite also leads to a 1000x, or more, increase in surface density at inner radii for high C/O ratios when compared to the Solar model, which is a result we share with \citet{Bond2010}. Systems with C/O=1 or greater host inner regions with nearly 100\% carbon compositions in conjunction with high surface densities. These systems simultaneously host regions with a rich spectrum of shared carbide-silicate compositions. Our results show a large overlap between the habitable zone and high-carbon regions for Solar-like stars with high C/O ratios.

\begin{acknowledgments}
Computer support was provided by UNLV’s National Supercomputing Center. CJS and JHS acknowledge support from the NSF grant AST-1910955. ML acknowledges support from the National Natural Science Foundation of China (NSFC) grant 12203018, a grant from the Siping Bureau of Science and Technology (2023071), and a grant from Jilin Normal University. SH acknowledges support from UTK Sisk endowed professorship. ZZ acknowledges support from NASA award 80NSSC22K1413.
\end{acknowledgments}

%






\appendix

\section{Additional Figures}
\label{sec:Appendix}

\begin{center}
\begin{figure*}[ht]
    \centering
	\includegraphics[width=0.48\textwidth]{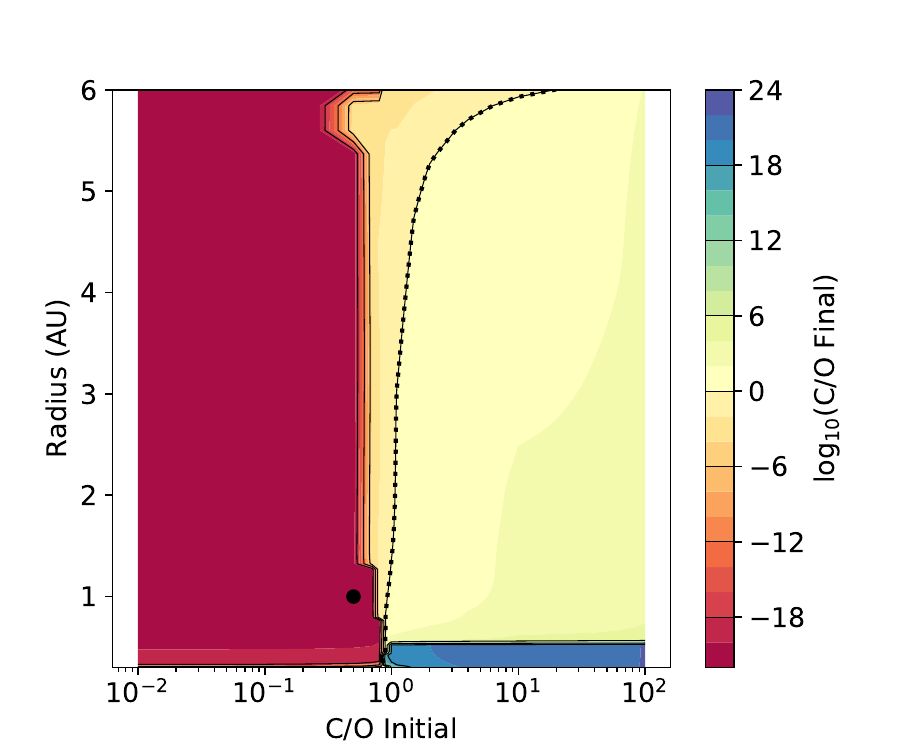}
	\includegraphics[width=0.48\textwidth]{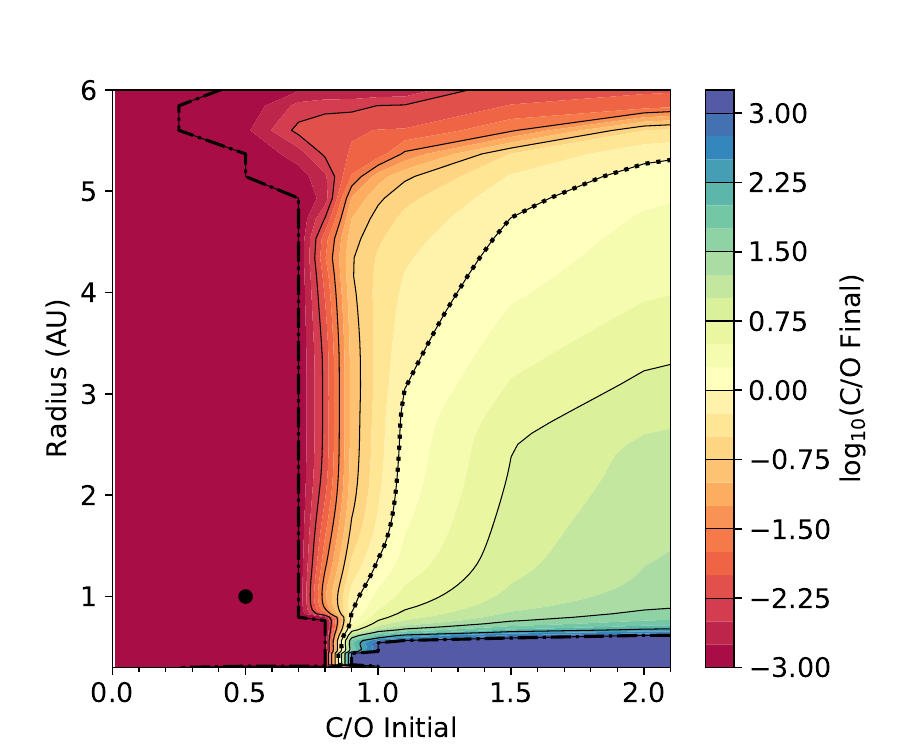}
    \caption{The same graphic as Figure \ref{fig:Frac3D} but here the amount of initial oxygen is held constant at the Solar value (i.e. w/Oc) while the initial C/O ratio varies.}
    \label{fig:Frac3DOc}
\end{figure*}
\end{center}
\begin{figure*}
	\includegraphics[width=0.96\textwidth]{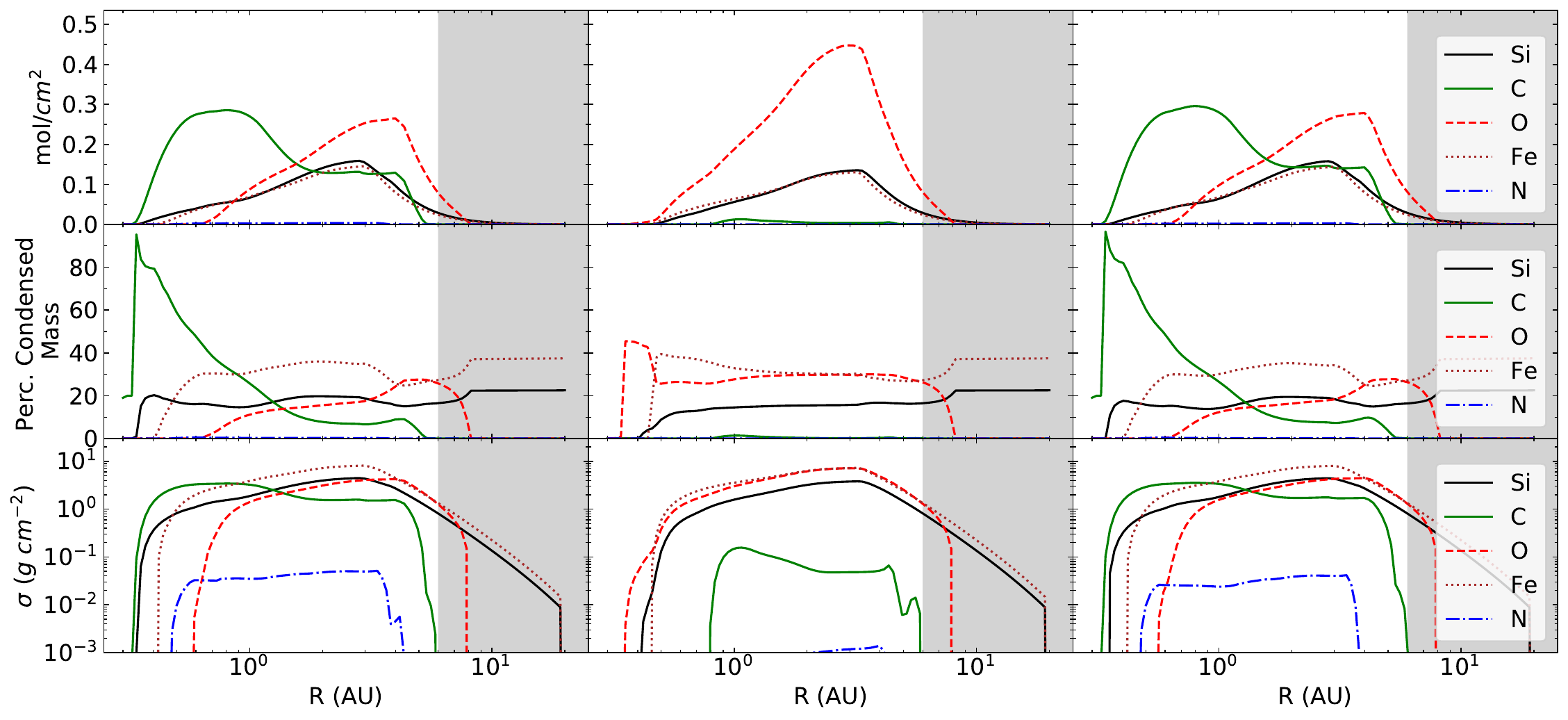}
    \caption{The solid surface density profiles of five select elements in three different simulations which hold either oxygen or carbon constant: (Left) C/O=1 w/Cc, (Middle) C/O=0.8 w/Oc, and (Right) C/O=1 w/Oc. Compare the corresponding C/O ratio here to the ``constant sum'' simulations in Figure \ref{fig:EleSig1}.}
    \label{fig:EleSigConst}
\end{figure*}

\newpage

\bibliography{sample631}{}

\begin{thebibliography}{}
\expandafter\ifx\csname natexlab\endcsname\relax\def\natexlab#1{#1}\fi
\providecommand{\url}[1]{\href{#1}{#1}}
\providecommand{\dodoi}[1]{doi:~\href{http://doi.org/#1}{\nolinkurl{#1}}}
\providecommand{\doeprint}[1]{\href{http://ascl.net/#1}{\nolinkurl{http://ascl.net/#1}}}
\providecommand{\doarXiv}[1]{\href{https://arxiv.org/abs/#1}{\nolinkurl{https://arxiv.org/abs/#1}}}

\bibitem[{{All{\`e}gre} {et~al.}(2001){All{\`e}gre}, {Manh{\`e}s}, \& {Lewin}}]{Allegre2001}
{All{\`e}gre}, C., {Manh{\`e}s}, G., \& {Lewin}, {\'E}. 2001, Earth and Planetary Science Letters, 185, 49, \dodoi{10.1016/S0012-821X(00)00359-9}

\bibitem[{{Allen-Sutter} {et~al.}(2020){Allen-Sutter}, {Garhart}, {Leinenweber}, {Prakapenka}, {Greenberg}, \& {Shim}}]{Allen2020}
{Allen-Sutter}, H., {Garhart}, E., {Leinenweber}, K., {et~al.} 2020, \psj, 1, 39, \dodoi{10.3847/PSJ/abaa3e}

\bibitem[{{Allibert} {et~al.}(2021){Allibert}, {Charnoz}, {Siebert}, {Jacobson}, \& {Raymond}}]{Allibert2021}
{Allibert}, L., {Charnoz}, S., {Siebert}, J., {Jacobson}, S.~A., \& {Raymond}, S.~N. 2021, \icarus, 363, 114412, \dodoi{10.1016/j.icarus.2021.114412}

\bibitem[{{Bond} {et~al.}(2010a){Bond}, {Lauretta}, \& {O'Brien}}]{Bond2010a}
{Bond}, J.~C., {Lauretta}, D.~S., \& {O'Brien}, D.~P. 2010a, \icarus, 205, 321, \dodoi{10.1016/j.icarus.2009.07.037}

\bibitem[{Bond {et~al.}(2010b)Bond, O’Brien, \& Lauretta}]{Bond2010}
Bond, J.~C., O’Brien, D.~P., \& Lauretta, D.~S. 2010b, The Astrophysical Journal, 715, 1050–1070, \dodoi{10.1088/0004-637x/715/2/1050}

\bibitem[{{Brewer} \& {Fischer}(2016)}]{Brewer2016}
{Brewer}, J.~M., \& {Fischer}, D.~A. 2016, \apj, 831, 20, \dodoi{10.3847/0004-637X/831/1/20}

\bibitem[{{Campbell} \& {St C. O'Neill}(2012)}]{Campbell2012}
{Campbell}, I.~H., \& {St C. O'Neill}, H. 2012, \nat, 483, 553, \dodoi{10.1038/nature10901}

\bibitem[{Cassen(1996)}]{Cassen1996}
Cassen, P. 1996, Meteoritics \& Planetary Science, 31, 793, \dodoi{10.1111/j.1945-5100.1996.tb02114.x}

\bibitem[{{Chambers}(2009)}]{Chambers2009}
{Chambers}, J.~E. 2009, \apj, 705, 1206, \dodoi{10.1088/0004-637X/705/2/1206}

\bibitem[{{Childs} {et~al.}(2023){Childs}, {Shakespeare}, {Rice}, {Yang}, \& {Steffen}}]{Childs2023}
{Childs}, A.~C., {Shakespeare}, C., {Rice}, D.~R., {Yang}, C.-C., \& {Steffen}, J.~H. 2023, \mnras, 524, 3749, \dodoi{10.1093/mnras/stad2110}

\bibitem[{{Gail} \& {Sedlmayr}(1986)}]{Gail1986}
{Gail}, H.~P., \& {Sedlmayr}, E. 1986, \aap, 166, 225

\bibitem[{{Hayashi}(1981)}]{Hayashi1981}
{Hayashi}, C. 1981, in Fundamental Problems in the Theory of Stellar Evolution, ed. D.~{Sugimoto}, D.~Q. {Lamb}, \& D.~N. {Schramm}, Vol.~93, 113--126

\bibitem[{{Hersant} {et~al.}(2001){Hersant}, {Gautier}, \& {Hur{\'e}}}]{Hersant2001}
{Hersant}, F., {Gautier}, D., \& {Hur{\'e}}, J.-M. 2001, \apj, 554, 391, \dodoi{10.1086/321355}

\bibitem[{{Hinkel} {et~al.}(2014){Hinkel}, {Timmes}, {Young}, {Pagano}, \& {Turnbull}}]{Hinkel2014}
{Hinkel}, N.~R., {Timmes}, F.~X., {Young}, P.~A., {Pagano}, M.~D., \& {Turnbull}, M.~C. 2014, \aj, 148, 54, \dodoi{10.1088/0004-6256/148/3/54}

\bibitem[{{Jorge} {et~al.}(2022){Jorge}, {Kamp}, {Waters}, {Woitke}, \& {Spaargaren}}]{Jorge2022}
{Jorge}, D.~M., {Kamp}, I.~E.~E., {Waters}, L.~B.~F.~M., {Woitke}, P., \& {Spaargaren}, R.~J. 2022, \aap, 660, A85, \dodoi{10.1051/0004-6361/202142738}

\bibitem[{{Kuchner} \& {Seager}(2005)}]{Kuchner2005}
{Kuchner}, M.~J., \& {Seager}, S. 2005, arXiv e-prints, astro, \dodoi{10.48550/arXiv.astro-ph/0504214}

\bibitem[{Li {et~al.}(2020)Li, Huang, Petaev, Zhu, \& Steffen}]{Li2020}
Li, M., Huang, S., Petaev, M.~I., Zhu, Z., \& Steffen, J.~H. 2020, \mnras, 495, 2543, \dodoi{10.1093/mnras/staa1149}

\bibitem[{{Lodders}(2003)}]{Lodders2003}
{Lodders}, K. 2003, \apj, 591, 1220, \dodoi{10.1086/375492}

\bibitem[{{Lodders}(2004)}]{Lodders2004}
---. 2004, \apj, 611, 587, \dodoi{10.1086/421970}

\bibitem[{Moriarty {et~al.}(2014)Moriarty, Madhusudhan, \& Fischer}]{Mori2014}
Moriarty, J., Madhusudhan, N., \& Fischer, D. 2014, The Astrophysical Journal, 787, 81, \dodoi{10.1088/0004-637x/787/1/81}

\bibitem[{{O'Brien} {et~al.}(2006){O'Brien}, {Morbidelli}, \& {Levison}}]{OBrien2006}
{O'Brien}, D.~P., {Morbidelli}, A., \& {Levison}, H.~F. 2006, \icarus, 184, 39, \dodoi{10.1016/j.icarus.2006.04.005}

\bibitem[{Petaev(2009)}]{Petaev2009}
Petaev, M.~I. 2009, Calphad, 33, 317, \dodoi{10.1016/j.calphad.2008.12.001}

\bibitem[{{Woitke} {et~al.}(2018){Woitke}, {Helling}, {Hunter}, {Millard}, {Turner}, {Worters}, {Blecic}, \& {Stock}}]{Woitke2018}
{Woitke}, P., {Helling}, C., {Hunter}, G.~H., {et~al.} 2018, \aap, 614, A1, \dodoi{10.1051/0004-6361/201732193}

\end{thebibliography}
\bibliographystyle{aasjournal}



\end{document}